\documentclass[letter,oneside,9pt,twocolumn,article]{aiaa}
\usepackage{rotating}
\usepackage{bm}
\usepackage{flushend}
\usepackage[top=0.5in, bottom=1.0in, left=0.7in, right=0.7in]{geometry}
\usepackage[bf,footnotesize]{subfigure}
\usepackage{adjustbox}
\usepackage{tabularx}
\usepackage{multirow}
\usepackage{multicol}
\usepackage{indentfirst} 
\usepackage[hidelinks]{hyperref}
\usepackage{float}
\usepackage{xcolor}
\usepackage[super,comma,sort,compress]{natbib}

\usepackage{mwe}
\usepackage[markcase=noupper]{scrlayer-scrpage}
\usepackage{caption} 

\ihead*{Submitted for publication in {\it Physics of Plasmas}}
\ohead*{\pagemark}
\ifoot*{}
\ofoot*{\rm Unrestricted content}
\newcommand{\alb}{\vspace{0.1cm}\\} 

\renewcommand{\fontsizetable}{\footnotesize\scalefont{1.0}}

\author{
Bernard Parent\thanks{Associate Professor, bparent@arizona.edu.} ~and Brendan Perry\thanks{Graduate student} 
\\[0.3em] \it University of Arizona, Tucson, AZ 85721, USA.
}

\title{
Birth-Potential Multigroup Fluid Model for Ballistic Electrons: Breakdown and Cathode Sheath Regimes
}

\abstract{ 
Fluid models of discharges fail where electron runaway dominates. On the left branch of the Paschen curve and in high-voltage cathode sheaths, the electron population splits into a thermal bulk and a ballistic beam that no single electron fluid can represent. We present a multigroup fluid model that divides ballistic electrons into groups indexed by their birth potential: the electrostatic potential at which they were liberated, offset by the forward kinetic energy they carried. Because the birth potential is a constant of ballistic motion, the electric field never moves electrons between groups and only collisions do. Each group requires a single continuity equation because its velocity follows algebraically from the local potential, and the number of groups is set by the energy resolution alone independently of the mesh. A runaway criterion based on the local field assigns each newly liberated electron to the bulk or a ballistic group, so the model applies even where no sheath exists. Benchmarked against a kinetic solver in argon, it reproduces the left branch of the Paschen curve---which local-field and local-energy models miss or misplace---and tracks the current--voltage characteristic of a cathode sheath to within a few percent at reduced fields exceeding $10^4$~Td, where the local-energy model overpredicts the current up to thirtyfold. Since as few as 10 to 30 groups suffice, depending on the field, the approach brings near-kinetic accuracy to multi-species computational fluid dynamics solvers at the cost of that many additional species, without added velocity-space dimensions or particle noise.
}

\nomenclature{
\begin{nomenclaturelist}{Roman symbols}
\item[$b$]{index of the parent ballistic group in a collision}
\item[$C_{\rm Ar^+}$]{charge of an argon ion, C}
\item[$C_{\rm e}$]{charge of the electron, C}
\item[$d$]{gap between the electrodes (breakdown case) or thickness of the cathode sheath (sheath case), m}
\item[$D_{\rm e}$]{Einstein diffusion coefficient of the bulk electrons, m$^2$/s}
\item[$e_{\rm e}$]{specific energy of the bulk electrons, J/kg}
\item[$E$]{electric field, V/m}
\item[$E_{\rm crit}$]{critical runaway field, the maximum of the friction curve, V/m}
\item[$E_\eta$]{component of the electric field normal to a boundary, V/m}
\item[$E^\star$]{reduced electric field, $|E|/N$, V$\cdot$m$^2$ or Td}
\item[$f_j$]{partition coefficient, share of a volumetrically liberated electron assigned to group $j$}
\item[$f_0$]{share assigned to the bulk group, $1-f_{\rm run}$}
\item[$f_{\rm run}$]{runaway fraction, share of a liberated electron the field carries away}
\item[$g_j$]{destination weight, share of a birth at $\phi_{\rm birth}$ received by group $j$}
\item[$H$]{Heaviside step function}
\item[$h_{\rm e}$]{specific enthalpy of the bulk electrons, J/kg}
\item[$j$]{electron group index, 0 for the bulk group}
\item[$j'$]{index of the group receiving the secondary electron of an ionizing collision}
\item[$j_{\rm A}$]{index of the group whose window brackets a given birth potential}
\item[$J_{\rm e}$]{electron current density, summed over the bulk and ballistic groups, A/m$^2$}
\item[$k$]{index of an inelastic collision channel, ionization included; $k={\rm iz}$ denotes the ionization channel}
\item[$k_1$]{electron-impact ionization rate coefficient, m$^3$/s}
\item[$k_2$]{ion-impact ionization rate coefficient, m$^3$/s}
\item[$k_{\rm B}$]{Boltzmann constant, J/K}
\item[$k^{\rm casc}_{j \leftarrow b,k}$]{degradation cascade rate coefficient, from group $b$ to group $j$ through channel $k$, m$^3$/s}
\item[$k^{\rm run}_{j \leftarrow 0}$]{thermal runaway rate coefficient, m$^3$/s}
\item[$k^{\rm scat}_{j \leftarrow b}$]{elastic scattering sink rate coefficient, m$^3$/s}
\item[$k^{\rm scat}_{j \leftarrow b,k}$]{inelastic scattering sink rate coefficient of channel $k$, m$^3$/s}
\item[$k^{\rm turn}_b$]{turning point rate coefficient, 1/s}
\item[$m_{\rm Ar^+}$]{mass of an argon ion, kg}
\item[$m_{\rm e}$]{mass of the electron, kg}
\item[$m_{\rm Ar}$]{mass of an argon atom, kg}
\item[$n_{\rm b}$]{number of ballistic electron groups}
\item[$N$]{total gas number density, m$^{-3}$}
\item[$N_{\rm Ar}$]{number density of argon atoms, m$^{-3}$}
\item[$N_{\rm Ar^+}$]{number density of argon ions, m$^{-3}$}
\item[$N_{\rm e}$]{total electron number density, summed over the bulk and ballistic groups, m$^{-3}$}
\item[$N_{{\rm e}_j}$]{number density of electron group $j$, m$^{-3}$ or cm$^{-3}$}
\item[$N_{{\rm e}_0}^{\rm nw}$]{bulk electron number density at the node adjacent to an absorbing wall, m$^{-3}$}
\item[$P$]{gas pressure, Pa (Torr in the product $Pd$)}
\item[$P_{\rm Ar^+}$]{partial pressure of the argon ions, Pa}
\item[$P_{\rm e}$]{partial pressure of the bulk electrons, Pa}
\item[$\overline{q_{\rm e}}$]{mean thermal speed, averaged over the bulk electrons alone, m/s}
\item[$Q_{{\rm e}_0}^{\rm beam}$]{energy delivered to the bulk group by the ballistic groups, W/m$^3$}
\item[$Q_{{\rm e}_0}^{\rm born}$]{energy delivered to the bulk group by newly created electrons, W/m$^3$}
\item[$Q_{{\rm e}_0}^{\rm relax}$]{energy exchanged with heavy particles by collisions, W/m$^3$}
\item[$Q_{{\rm e}_0}^{\rm run}$]{energy leaving the bulk group with the mass that runs away, W/m$^3$}
\item[$r_j$, $r$]{ratio of the width of the adjacent lower-energy window $j-1$ to that of group $j$, the subscript dropped where no confusion arises}
\item[$s$]{position of a birth within its window, measured from the upper bound}
\item[$t$]{time, s}
\item[$T$]{gas temperature, K}
\item[$T_{\rm born}$]{temperature at which electrons are created in the volume, K}
\item[$T_{\rm e}$]{temperature of the bulk electrons, K or eV}
\item[$T_{\rm ref}$]{reference temperature of the elastic relaxation term, K}
\item[$u$]{birth potential in units of the level width, measured from the midpoint}
\item[$V_{\rm Ar^+}$]{macroscopic velocity of the argon ions, m/s}
\item[$V_{\rm c}$]{cathode-fall voltage, V}
\item[$V_{\rm e}$]{speed of an individual electron, m/s}
\item[$V_{\rm e}^x$]{$x$ component of the velocity of an individual electron, m/s}
\item[$V_{{\rm e}_j}$]{macroscopic velocity of electron group $j$, m/s}
\item[$W_{\rm Ar^+}$]{chemical source of the argon ions, kg/(m$^3\cdot$s)}
\item[$W_{{\rm e}_j}$]{chemical source of electron group $j$, kg/(m$^3\cdot$s)}
\item[$x$]{distance from the cathode, m}
\item[$y$]{local potential in units of the level width, measured from the midpoint}
\item[$z$]{distance beyond the join of the birth deposit, in units of the level width}
  \end{nomenclaturelist}
  \begin{nomenclaturelist}{Greek symbols}
\item[$\chi_0$]{fraction of post-collision directions from which a thermal electron can be promptly re-collimated, set to $0.5$}
\item[$\chi_{b,k}$]{fraction of its forward momentum an electron of group $b$ retains in a collision of channel $k$}
\item[$\gamma_{\rm e}$]{secondary electron emission coefficient, electrons emitted per incident ion}
\item[$\delta_{j0}$]{Kronecker delta, unity for the bulk group and zero otherwise}
\item[$\Delta E$]{transition field width of the runaway partition, V/m}
\item[$\Delta x$]{width of a computational cell, m}
\item[$\Delta \varepsilon_k$]{threshold energy of inelastic channel $k$, J or eV}
\item[$\Delta \phi_j$]{width in potential of the window of group $j$, V}
\item[$\epsilon_0$]{permittivity of free space, F/m}
\item[$\varepsilon$]{kinetic energy of an electron, J or eV}
\item[$\varepsilon_{+x}$]{forward energy of an electron, its energy of motion along $+x$, J or eV}
\item[$\varepsilon_{\rm born}$]{energy of an electron created in the volume, $\tfrac{3}{2} k_{\rm B} T_{\rm born}$, J or eV}
\item[$\varepsilon_{\rm emit}$]{energy of an electron emitted from the cathode surface, J or eV}
\item[$\varepsilon_{{\rm e}_j}$]{kinetic energy of electron group $j$, J or eV}
\item[$\langle \varepsilon \rangle$]{mean energy per electron, averaged over all electrons, bulk and ballistic, J or eV}
\item[$\overline{\varepsilon_{\rm e}}$]{mean energy per electron, averaged over the bulk electrons alone, $\tfrac{3}{2} k_{\rm B} T_{\rm e}$, J or eV}
\item[$\varepsilon^{\rm lo}_j$, $\varepsilon^{\rm hi}_j$]{energy bounds of the window of group $j$, J or eV}
\item[$\zeta$]{straggling parameter, the transition field width relative to the critical field, set to $0.8$}
\item[$\eta$]{spatial coordinate normal to a boundary, positive toward the plasma, m}
\item[$\theta$]{angle between an electron's direction of motion and the $+x$ axis, equal to the collisional deflection angle for a beam electron}
\item[$\kappa_{\rm e}$]{thermal conductivity of the bulk electrons, W/(m$\cdot$K)}
\item[$\ln \Lambda$]{Coulomb logarithm}
\item[$\mu_{\rm e}$]{mobility of the bulk electrons, m$^2$/(V$\cdot$s)}
\item[$\mu_{\rm Ar^+}$]{mobility of the argon ions, m$^2$/(V$\cdot$s)}
\item[$\mu_{\rm e}^\star$]{reduced electron mobility, $\mu_{\rm e} N$, 1/(m$\cdot$V$\cdot$s)}
\item[$\nu^{\rm casc}_{b,k}$]{collision frequency of channel $k$ for group $b$, 1/s}
\item[$\nu^{\rm scat}_b$]{de-beaming frequency of group $b$, 1/s}
\item[$\xi_j$]{displacement of the mean group potential $\overline{\phi}_j$ from the midpoint of its window, in units of the level width}
\item[$\rho_{\rm Ar^+}$]{mass density of the argon ions, kg/m$^3$}
\item[$\rho_{{\rm e}_j}$]{mass density of electron group $j$, kg/m$^3$}
\item[$\sigma_{\rm deb}$]{de-beaming cross-section, the sum of the elastic and deflected inelastic contributions, m$^2$}
\item[$\sigma_{\rm el}$]{momentum transfer cross-section for elastic collisions, m$^2$}
\item[$\sigma_{\rm iz}$]{cross-section for electron-impact ionization, m$^2$}
\item[$\sigma_k$]{cross-section of inelastic channel $k$, ionization included, m$^2$}
\item[$\phi$]{electrostatic potential, V}
\item[$\phi_{\rm birth}$]{birth potential of an electron, V}
\item[$\phi^{\rm birth}_0$]{birth potential of an electron leaving the bulk group by thermal runaway, V}
\item[$\phi^{\rm birth}_{b,k}$]{birth potential of an electron of group $b$ after a collision of channel $k$, V}
\item[$\overline{\phi}_j$]{mean group potential of group $j$, from which its energy is measured, V}
\item[$\phi_{\rm cathode}$]{potential of the cathode, V}
\item[$\phi_{\rm emit}$]{birth potential of the electrons emitted from the cathode surface, V}
\item[$\phi_{\rm gauge}$]{potential to which the zero of the energy axis is mapped, V}
\item[$\phi^{\rm lo}_j$, $\phi^{\rm hi}_j$]{lower and upper bounds of the window of group $j$, V}
\item[$\phi^{\rm mid}_j$]{midpoint of the window of group $j$, V}
  \end{nomenclaturelist}
}

\begin{document}
\maketitle

\makenomenclature

\section{Introduction}

\dropword
Fluid models of plasmas struggle to capture the left side of the Paschen curve and high-voltage cathode sheaths, regimes where electron transport becomes primarily ballistic \cite{jpd:2009:derzsi}. In these regions, the electron population splits into a cold thermal bulk and a high-energy beam, creating a strongly bimodal electron energy distribution function (EEDF). Traditional single-fluid approaches, which average properties over the entire population, fail to capture this severe departure from local equilibrium.

To resolve this bimodal EEDF, most studies rely on kinetic or hybrid frameworks. Fully kinetic models capture non-local behavior and secondary electron dynamics faithfully, but their large computational cost \cite{Becker_2017, Schulze_2011, Wilczek_2015, Kortshagen_1996, Kortshagen_1995, Quraishi_2020, Noesges_2023, Derzsi_2019} often necessitates compromises that can lead to under-resolved phenomena and, in particle methods, severe numerical noise. Hybrid Monte Carlo-fluid models \cite{Sommerer1992, Bogaerts_1999, Sommerer_1993, Eremin_2015, Wen_2022} mitigate this by tracking only the fast electrons kinetically, but they remain computationally expensive and frequently suffer from numerical coupling issues between the kinetic and fluid populations.

As a more efficient alternative, macroscopic fluid frameworks have been continuously refined \cite{Zhang_2024, Zhao_2018, Young_1993}, though widely used approximations often fail in nonlocal cathode regions \cite{psst:2024:zhou} or rely on analytic sources limited to steady discharges with well-defined sheaths \cite{pop:2012b:rafatov, psst:2020:yao}. To handle ballistic phenomena, previous models have partitioned electrons into beam and bulk groups, but often relied on restrictive assumptions such as a Local Field Approximation (LFA) for the bulk, artificial thermalization cutoffs, or steady-state limitations \cite{pra:1989:gottscho, Belenguer1990, Hashiguchi_1993, Chanrion_2014}. Finer multibeam discretizations---whether treating discrete beams originating at birth points \cite{zp:1962:muller, pra:1987:phelps, apl:1990:pak} or binning forward-directed flux by kinetic energy \cite{jpd:1987:carman, pra:1990:surendra}---have also faced challenges. These approaches generally tie the spatial mesh to a monotone potential profile, prescribe a one-way flow of mass into the bulk, or neglect scattering entirely, which leads to excessive penetration of the fast electrons \cite{pra:1990:surendra}. Consequently, these models have not been shown to reproduce kinetic predictions on the left side of the Paschen curve, where no sheath exists and ballistic electrons originate in the volume.

This paper introduces a multigroup electron fluid model that partitions electrons into a thermal bulk fluid and $n_{\rm b}$ discrete ballistic inertial groups, indexed by their electrostatic birth potential. Building on nonlocal approaches that express the electron distribution in terms of total energy \cite{pr:1954:bernstein, psst:1995:tsendin, Kortshagen_1996, pra:1992:kolobov}, we use this birth potential as an explicit group index independent of the spatial mesh. Electrons liberated within the same birth potential interval join the corresponding group, regardless of their birth location. To our knowledge, this specific use of birth potential is novel. Previous schemes tracked total energy only implicitly via potential-conforming spatial meshes \cite{jpd:1987:carman, pra:1990:surendra}, launched separate beams from each birth location \cite{zp:1962:muller, pra:1987:phelps, apl:1990:pak, jpd:1991:kulikovsky}, or treated it as a continuous coordinate in analytic kinetic solutions \cite{pra:1992:kolobov}. In contrast, our group count is determined solely by the desired energy resolution, decoupling it from the number of grid cells or source points. Unlike multigroup models that bin electrons by kinetic energy \cite{jetp:2007:babich, ppr:2023:bochkov, ppr:2023b:bochkov, ppr:2024:bochkov}, where each group requires both a density and a flux equation and the field advects electrons across group boundaries, the field never moves electrons from one group to another here; only collisions do.

This decoupling keeps the approach computationally tractable for mesh refinement and multidimensional expansion. While traditional beam-per-location schemes scale poorly with cell count and face prohibitive renormalization challenges in two- and three-dimensional sheaths \cite{jpd:1991:kulikovsky}, our model's cost scales linearly as the number of cells multiplied by a fixed number of groups. Although demonstrated here in 1D, this scalability is important for multidimensional sheath modeling. Computationally, each group is governed by a single continuity equation on the bulk grid. Because a group's energy is measured from a mean group potential spanning its interval, its velocity is derived algebraically from the local potential, eliminating the need for separate momentum or energy equations. Whether a newly liberated electron joins the bulk or a ballistic group is decided by the local electric field, so the model applies even where no sheath exists, as on the left branch of the Paschen curve. The resulting model captures the bimodal EEDF with near-kinetic fidelity across both breakdown and cathode-sheath regimes.

The paper is structured as follows: we first detail the transport equations for the ballistic and bulk electron groups and for the heavy particles. We then validate the proposed multigroup model against standard fluid and kinetic models. Our benchmarks focus on breakdown voltages along both branches of the Paschen curve and on a high-voltage cathode sheath, two regimes where standard fluid models fail to predict the current-voltage relationship accurately.

\section{Ballistic Electron Groups}
\label{sec:ballistic}

The electron population is partitioned into $n_{\rm b}+1$ groups: a bulk group, with index 0, assumed to be in local equilibrium and without inertia, and $n_{\rm b}$ ballistic groups that retain their inertia. Groups $1$ to $n_{\rm b}-1$ are the volumetric ballistic groups, which receive the electrons liberated in the gas volume wherever the field can run them away, ordered by energy with group $1$ the least energetic. Group $n_{\rm b}$ is reserved for the secondary electrons emitted from the cathode surface. All ballistic groups travel in the $+x$ direction, with $x$ the distance from the cathode, and any electron that would be accelerated in the opposite direction is reassigned to the bulk group; the model is therefore confined to configurations in which the left electrode serves as the cathode. The ballistic groups are described in this section and the bulk group in Section~\ref{sec:bulk}.

What distinguishes one ballistic group from another is the potential at which its members were liberated. A ballistic electron conserves its total energy $\varepsilon + C_{\rm e} \phi$, with $C_{\rm e}$ the electron charge, the sum of its kinetic energy $\varepsilon$ and its potential energy in the electrostatic potential $\phi$, so it retains whatever it held at birth until a collision occurs. As in the nonlocal kinetic approach \cite{pr:1954:bernstein, psst:1995:tsendin, Kortshagen_1996}, we exploit this conservation of total energy between collisions, here to label the ballistic groups. The label is the birth potential $\phi_{\rm birth}$, the difference between the potential $\phi$ at which an electron is delivered to a ballistic group and the \emph{forward} energy $\varepsilon_{+x}$ it holds prior to joining the group:
\begin{equation}
\phi_{\rm birth} = \phi - \frac{\varepsilon_{+x}}{|C_{\rm e}|}
\label{eqn:phibirth}
\end{equation}
in which $\varepsilon_{+x} = \tfrac{1}{2} m_{\rm e} \left[\max\left(0, V_{\rm e}^x\right)\right]^2$, with $m_{\rm e}$ the electron mass and $V_{\rm e}^x$ the $x$ component of the electron velocity: the kinetic energy of motion along the beam direction alone, which is what a group can carry and what the field can return. Since $C_{\rm e}<0$, the birth potential of an electron moving along $+x$ is its total energy divided by $C_{\rm e}$, and it therefore remains constant between collisions. A ballistic group is a beam of a single speed in a single direction, so for any electron a group holds, the forward energy is the whole of its kinetic energy. The distinction matters only at the moment of delivery, where an electron may arrive with its energy pointed anywhere.

The forward energy at delivery is set by the process that delivers the electron, and there are four. One liberated in the gas volume is created at the birth temperature $T_{\rm born}$ (1600~K here), whose energy is thermal and hence isotropic. It carries no net forward motion, and whatever forward energy it holds individually, a fraction of $\tfrac{3}{2}k_{\rm B}T_{\rm born}\approx 0.2$~eV with $k_{\rm B}$ the Boltzmann constant, is negligible against the width of a level, so it is given $\varepsilon_{+x} = 0$ and placed at $\phi_{\rm birth} = \phi$. One emitted from the cathode surface is directed into the gap and carries $\varepsilon_{+x} = \varepsilon_{\rm emit}$, the mean initial kinetic energy of the secondaries as they leave the surface. One passed from a group to another by a collision carries what is left of the energy it held in the group it came from, provided the collision left it travelling forward; if not, it is re-partitioned as though newly liberated, with $\varepsilon_{+x} = 0$ (Section~\ref{sec:cascade}). And one that runs away from the bulk group carries approximately its mean thermal energy forward, $\varepsilon_{+x} \approx \overline{\varepsilon_{\rm e}} = \tfrac{3}{2}k_{\rm B}T_{\rm e}$, with $T_{\rm e}$ the bulk electron temperature (Section~\ref{sec:thermalrunaway}).

The birth-potential axis is divided into windows, one per volumetric group. Their boundaries are supplied in electron-volts as an increasing list of $n_{\rm b}$ levels:
\begin{equation}
\varepsilon^{\rm lo}_1 < \varepsilon^{\rm lo}_2 < \cdots < \varepsilon^{\rm lo}_{n_{\rm b}-1} < \varepsilon^{\rm hi}_{n_{\rm b}-1}
\end{equation}
with group $j$ spanning $[\varepsilon^{\rm lo}_j, \varepsilon^{\rm hi}_j)$ and $\varepsilon^{\rm hi}_j = \varepsilon^{\rm lo}_{j+1}$ for $j < n_{\rm b}-1$. Being prescribed rather than uniformly subdivided, the levels may be concentrated across the ionization onset and the peak of the electron-impact ionization cross-section $\sigma_{\rm iz}$. The list is carried onto the potential axis through the remap
\begin{equation}
\phi(\varepsilon) = \phi_{\rm gauge} - \frac{\varepsilon}{|C_{\rm e}|}
\label{eqn:ev2volt}
\end{equation}
in which the gauge potential $\phi_{\rm gauge}$ must be at least the larger of the potential drop across the cathode sheath and the potential difference between the electrodes. Overestimating it does not make the solution erroneous but increases the computational cost for an equivalent numerical error. Because the map is decreasing, the upper energy bound of a group becomes its lower bound in potential:
\begin{equation}
\phi^{\rm lo}_j = \phi\big(\varepsilon^{\rm hi}_j\big), \qquad \phi^{\rm hi}_j = \phi\big(\varepsilon^{\rm lo}_j\big)
\label{eqn:philo}
\end{equation}
Because an electron born at a lower potential has more potential to climb, it is the more energetic at any given location, and the window of group $j$ lies below that of the less energetic group $j-1$. Each window is also assigned its midpoint,
\begin{equation}
\phi^{\rm mid}_j = \tfrac{1}{2}\left( \phi^{\rm lo}_j + \phi^{\rm hi}_j \right) \quad \text{for } j \in \{1, \ldots, n_{\rm b}-1\}
\label{eqn:epsmid}
\end{equation}
and its width $\Delta \phi_j = \phi^{\rm hi}_j - \phi^{\rm lo}_j$. These four numbers per group follow from the level list and $\phi_{\rm gauge}$ alone and do not change during the computation. Group $j$ receives the electrons whose birth potential lies within its window, $\phi^{\rm lo}_j \le \phi_{\rm birth} < \phi^{\rm hi}_j$, wherever in the domain they are liberated; to keep the transfer continuous, each birth is shared linearly with the adjacent more energetic group (Section~\ref{sec:runaway}).

The emission group $j=n_{\rm b}$ takes no level from the list. Its members all leave the cathode at the same potential with the same forward energy, so by Eq.\,\eqref{eqn:phibirth} they share a single birth potential and its window is a point rather than an interval:
\begin{equation}
\phi^{\rm lo}_{n_{\rm b}} = \phi^{\rm hi}_{n_{\rm b}} = \phi^{\rm mid}_{n_{\rm b}} = \phi_{\rm emit}
\label{eqn:emitgroupwindow}
\end{equation}
with $\phi_{\rm emit} = \phi_{\rm cathode}-\varepsilon_{\rm emit}/|C_{\rm e}|$, $\phi_{\rm cathode}$ the potential of the cathode, and $\varepsilon_{\rm emit}$ set here to 1~eV.

The kinetic energy of group $j$ is determined by the potential its members have climbed since birth. It is measured from the mean group potential $\overline{\phi}_j$, a ramp across the window that replaces the midpoint to avoid spurious oscillations (Section~\ref{sec:centroid}); for the emission group, whose window is a point, $\overline{\phi}_{n_{\rm b}} = \phi_{\rm emit}$. Each group's kinetic energy is thus
\begin{equation}
\varepsilon_{{\rm e}_j}(\phi) = |C_{\rm e}| \left( \phi - \overline{\phi}_j(\phi) \right)
\label{eqn:eps_group}
\end{equation}
and because these particles move ballistically, their acquired energy is entirely kinetic and unidirectional. The velocity of each group therefore follows directly from the local potential:
\begin{equation}
V_{{\rm e}_j} = \sqrt{\frac{2 \varepsilon_{{\rm e}_j}}{m_{\rm e}}} \quad \text{for } j \in \{1, \ldots, n_{\rm b}\}
\label{eqn:V_group}
\end{equation}

These relationships establish two critical properties of the formulation. First, the group velocity $V_{{\rm e}_j}$ is independent of the group density $\rho_{{\rm e}_j}$, meaning that adding mass to a group neither dilutes its energy nor shifts the energy at which its collision cross-sections are evaluated. Second, because the exact acceleration and velocity are determined simply by knowing the local potential, there is no need to solve a separate momentum or energy transport equation for each group. Consequently, each ballistic group is governed solely by a single mass conservation equation:
\begin{equation}
\frac{\partial \rho_{{\rm e}_j}}{\partial t} + \frac{\partial }{\partial x} \left( \rho_{{\rm e}_j} V_{{\rm e}_j} \right) = W_{{\rm e}_j} \quad \text{for } j \in \{1, \ldots, n_{\rm b}\}
\label{eqn:masstransport_ej}
\end{equation}
with $t$ the time. The total electron density is the sum of the bulk electron density and the electron densities of all ballistic groups. The source $W_{{\rm e}_j}$ collects the mechanisms of the following subsections and is assembled in Section~\ref{sec:ballisticsource}.

\subsection{Runaway partition}
\label{sec:runaway}

This section derives the source partition coefficients $f_j$. When new electrons are liberated, these coefficients determine both whether they become ballistic and which specific group $j$ receives them.

An electron liberated at rest faces two competing forces. The electric field $E$ supplies energy to an electron of speed $V_{\rm e}$ at a rate $C_{\rm e} E V_{\rm e}$. Conversely, the argon gas, of number density $N_{\rm Ar}$, removes energy through inelastic collisions at a rate $N_{\rm Ar} \sum_k \sigma_k(\varepsilon) \Delta \varepsilon_k \, V_{\rm e}$, where $\sigma_k$ and $\Delta \varepsilon_k$ are the cross-section and threshold energy of inelastic channel $k$, ionization included. Expressed as an equivalent electric field, the maximum of this collisional loss rate defines the critical runaway field:
\begin{equation}
E_{\rm crit} = \frac{1}{|C_{\rm e}|} \max_{\varepsilon} \left[ N_{\rm Ar} \sum_k \sigma_k(\varepsilon)\, \Delta \varepsilon_k \right]
\label{eqn:Ecrit}
\end{equation}
For argon, this peak loss occurs at $\varepsilon = 75.0$~eV. By assuming the field supplies $C_{\rm e} E$ per unit path rather than the directional average $C_{\rm e} E \langle \cos \theta \rangle$, with $\theta$ the angle between the electron's direction of motion and the $+x$ axis, Eq.~\eqref{eqn:Ecrit} applies strictly to a perfectly collimated electron. It therefore serves as an absolute lower bound; below this field, no electrons can run away to join a ballistic group. It also counts only the threshold energy of each channel and not the energy carried off by the secondary electron in ionization; including the latter moves the peak to about 110~eV and raises $E_{\rm crit}/N$ from about 590 to 877~Td \cite{ppr:2025:bochkov}, a difference partly absorbed by the calibration of $\zeta$ below.

A deterministic step function at $E_{\rm crit}$ ignores collisional fluctuations. An electron may still fail to run away if its accumulated collisional losses exceed the energy excess $|C_{\rm e}| ( |E| - E_{\rm crit} )$ before it surpasses the friction peak. Because these collisions are a sequence of small, Poisson-distributed events, the failure probability decays exponentially with this excess. This yields the following expression for the runaway fraction $f_{\rm run}$:
\begin{equation}
f_{\rm run} = \begin{cases} 1 - \exp \left[ - \dfrac{|E| - E_{\rm crit}}{\Delta E} \right] & \text{for } -E > E_{\rm crit}  \\[3ex] 0 & \text{otherwise} \end{cases}
\label{eqn:frunstep}
\end{equation}
Here, the field's excess over $E_{\rm crit}$ is scaled by a transition width $\Delta E$, which represents the smearing effect of collisional straggling.

Equation~\eqref{eqn:frunstep} evaluates runaway using the local electric field. Physically, an electron must sustain this excess field over its entire climb to the friction peak---a potential drop of roughly 75~V in argon. The local formulation assumes the field remains nearly constant over this distance. This assumption holds precisely where the runaway partition matters most. In a high-voltage cathode sheath, this climb spans less than the first fifth of the sheath, a region where the strong field varies slowly. Likewise, during breakdown, the field is nearly uniform across the entire gap. The local form only overestimates runaway near the sheath edge if the total cathode fall voltage is small. However, very few volumetric electrons are produced in that region anyway. Tkachev and Yakovlenko \cite{tpl:2003:tkachev} argued that runaway is governed instead by the ratio of the gap to the Townsend multiplication length, a nonlocal criterion. Their criterion concerns the fraction of runaway electrons reaching the anode, which in a long gap is diluted by multiplication even when individual electrons run away; in the present model that dilution arises from the transport of the ballistic and bulk groups rather than from the partition itself.

The transition width $\Delta E$ is modeled as proportional to the critical field:
\begin{equation}
\Delta E = \zeta \, E_{\rm crit}
\label{eqn:deltaE}
\end{equation}
The straggling parameter is calibrated to $\zeta = 0.8$, the value that gives the closest agreement with the kinetic electron current over all the cathode-fall voltages of the sheath case in Section~\ref{sec:sheath}; the sensitivity of the solution to this choice is examined in Section~\ref{sec:sensitivity}. This constant is empirically chosen rather than derived from first principles. It may require adjustment for gases other than argon.

Liberated electrons that fail to run away immediately thermalize into the bulk group. The fraction assigned to the bulk is simply the complement of the runaway fraction:
\begin{equation}
f_0 = 1 - f_{\rm run} 
\label{eqn:f1}
\end{equation}
The runaway share must then be distributed among the eligible ballistic groups. Electrons can only join groups whose birth potentials do not exceed the potential at their point of liberation. The assignment is determined by the effective birth potential $\phi_{\rm birth}$ from Eq.\,\eqref{eqn:phibirth}. The distribution stencil is anchored to the group $j_{\rm A}$ whose bounds contain $\phi_{\rm birth}$:
\begin{equation}
j_{\rm A}(\phi_{\rm birth}) = j \ \text{ such that } \ \phi^{\rm lo}_j \le \phi_{\rm birth} < \phi^{\rm hi}_j
\label{eqn:jA}
\end{equation}
where $j$ ranges over the volumetric groups $[1..n_{\rm b}-1]$. The electron's relative position $s$ within this window is measured from the upper bound and normalized by the window's width:
\begin{equation}
s(\phi_{\rm birth}) = \frac{\phi^{\rm hi}_{j_{\rm A}} - \phi_{\rm birth}}{\phi^{\rm hi}_{j_{\rm A}} - \phi^{\rm lo}_{j_{\rm A}}} \in [0,1)
\label{eqn:sfrac}
\end{equation}
The runaway mass is then linearly partitioned between the anchor group $j_{\rm A}$ and the adjacent more energetic group $j_{\rm A}+1$. This yields the piecewise destination weights $g_j$:
\begin{equation}
g_j(\phi_{\rm birth}) = \begin{cases} 1 - s(\phi_{\rm birth}) & j = j_{\rm A}(\phi_{\rm birth}) \\[1ex] s(\phi_{\rm birth}) & j = j_{\rm A}(\phi_{\rm birth}) + 1 \\[1ex] 0 & \text{otherwise} \end{cases}
\label{eqn:kernel}
\end{equation}
The stencil restricts the upper index to $j_{\rm A}+1 \le n_{\rm b}-1$, as the reserved group $n_{\rm b}$ receives no volumetric mass. For births within the most energetic volumetric group, the boundary share $s$ is entirely retained by that group, ensuring $g_{n_{\rm b}-1} = 1$. This linear weighting guarantees that $\sum g_j = 1$ for any $s$, strictly conserving the volumetric source. It also ensures that only two weights are non-zero at any location, providing continuous mass transfer across group boundaries.

Finally, the absolute source partition coefficients $f_j$ are calculated by multiplying these destination weights by the runaway fraction, evaluated at $\phi_{\rm birth} = \phi(x)$:
\begin{equation}
f_j = f_{\rm run} \, g_j\big(\phi(x)\big)
\label{eqn:jstar}
\end{equation}
By definition, these coefficients satisfy $\sum_{j=1}^{n_{\rm b}-1} f_j = f_{\rm run}$. Combined with the bulk fraction, they guarantee that $f_0 + \sum_{j=1}^{n_{\rm b}-1} f_j = 1$. This ensures a rigorous and complete partition of every liberated electron.

\subsection{Mean group potential $\overline{\phi}_j$}
\label{sec:centroid}

The natural reference for the kinetic energy of group $j$ in Eq.\,\eqref{eqn:eps_group} is the window midpoint $\phi^{\rm mid}_j$. A static midpoint, however, gives an electron born at the lower bound an instantaneous kinetic energy of $|C_{\rm e}|\Delta\phi_j/2$. Across the group family, this introduces an unphysical staircase in group velocity and artificial steps in density at window boundaries, which are particularly pronounced on coarse energy grids. To ensure continuous acceleration and eliminate these numerical artifacts, we replace $\phi^{\rm mid}_j$ by a continuous mean group potential $\overline{\phi}_j(\phi)$ that ramps smoothly across the window:
\begin{equation}
\overline{\phi}_j(\phi) = \phi^{\rm mid}_j + \Delta \phi_j \, \xi_j(\phi) \quad \text{for } j \in \{1, \ldots, n_{\rm b}-1\}
\label{eqn:phibar}
\end{equation}
where $\xi_j(\phi)$ is a dimensionless offset. We define $\overline{\phi}_j(\phi)$ as the centroid of the birth potentials deposited into group $j$ up to the local potential $\phi$, assuming a birth density uniform in $\phi_{\rm birth}$:
\begin{equation}
\overline{\phi}_j(\phi) = \frac{\displaystyle \int_{-\infty}^{\phi} \phi_{\rm birth} \, g_j(\phi_{\rm birth}) \, {\rm d}\phi_{\rm birth}}{\displaystyle \int_{-\infty}^{\phi} g_j(\phi_{\rm birth}) \, {\rm d}\phi_{\rm birth}}
\label{eqn:centroid_def}
\end{equation}
evaluated using the destination weights $g_j(\phi_{\rm birth})$ defined in Eq.\,\eqref{eqn:kernel}. 

Physically, this definition ensures that a group appears at its lower bound with zero kinetic energy ($\overline{\phi}_j = \phi^{\rm lo}_j$, corresponding to $\xi_j = -\tfrac{1}{2}$). As $\phi$ rises across the group's collection window, $\overline{\phi}_j$ increases monotonically as newly liberated electrons are incorporated. Once $\phi$ exceeds the upper bound of the group's collection domain, no further volumetric mass can be added; $\overline{\phi}_j$ saturates at a constant value, and the group's energy subsequently increases purely ballistically with the electrostatic potential. 

Because $g_j(\phi_{\rm birth})$ varies linearly across adjacent group windows, Eq.\,\eqref{eqn:centroid_def} integrates analytically to piecewise rational functions. The final closed-form algebraic expressions for all branches of $\xi_j(\phi)$ are provided in Appendix~\ref{sec:appA} for direct implementation.

\subsection{Degradation cascade}
\label{sec:cascade}

The degradation cascade transfers mass between ballistic groups via inelastic collisions; rather than halting an electron entirely, these collisions extract energy, thereby demoting the electron to a lower energy group. 

The cascade is governed by a set of inelastic processes, indexed by $k$, encompassing both excitation and ionization. Each process contributes through its cross-section $\sigma_k(\varepsilon)$ and an energy loss $\Delta \varepsilon_k$ defined exclusively by its threshold. Elastic collisions are excluded from this framework, as their kinematic energy loss is too small to displace an electron by a resolvable fraction of an energy group spacing. For argon, we use the 2003 Hayashi database \cite{report:2003:hayashi} accessed via LXCat \cite{lxcat:hayashi}, which includes 25 excitation transitions and a single ionization channel at $\Delta \varepsilon_{\rm iz} = 15.76$~eV.

Following an inelastic collision of type $k$, a primary electron in group $b$ experiences an energy drop from $\varepsilon_{{\rm e}_b}(\phi)$ to $\varepsilon_{{\rm e}_b}(\phi) - \Delta \varepsilon_k$. To determine the degraded electron's new group assignment, we evaluate its updated birth potential. The share $\chi_{b,k}$ that survives the collision as a beam is taken to carry the whole of the remaining energy forward, so that $\varepsilon_{+x} = \varepsilon_{{\rm e}_b}(\phi) - \Delta \varepsilon_k$ for it; the complementary share, which does not, is removed by the scattering sink instead. Substituting the group energy $\varepsilon_{{\rm e}_b}$ from Eq.\,\eqref{eqn:eps_group} into the definition of the birth potential in Eq.\,\eqref{eqn:phibirth}, and accounting for the collision's energy loss, yields the effective birth potential of the post-collision electron:
\begin{equation}
\phi^{\rm birth}_{b,k}(\phi) = \overline{\phi}_b(\phi) + \frac{\Delta \varepsilon_k}{|C_{\rm e}|}
\label{eqn:phibcasc}
\end{equation}
Because the destination weight kernel $g_j$ in Eq.\,\eqref{eqn:kernel} is defined natively on the birth potential, evaluating $g_j\bigl(\phi^{\rm birth}_{b,k}\bigr)$ directly provides the fractional shares distributed to the destination groups $j$.

The fundamental frequency for a group $b$ electron to undergo an inelastic process $k$ is the standard collision frequency:
\begin{equation}
\nu^{\rm casc}_{b,k} = N_{\rm Ar} \, \sigma_k \! \left( \varepsilon_{{\rm e}_b} \right) V_{{\rm e}_b}
\label{eqn:nucasc}
\end{equation}
where $N_{\rm Ar}$ is the neutral number density and no fractional-loss weighting has yet been applied. However, not every inelastic collision preserves the electron's ballistic trajectory. A collision that extracts a large fraction of the primary's energy also scatters it through a large angle, depositing the emerging electron into the isotropic bulk population. Assuming a binary encounter between equal masses, the retained forward momentum is determined entirely by the fractional energy loss. An electron surrendering $\Delta \varepsilon_k / \varepsilon_{{\rm e}_b}$ emerges at an angle $\theta$ given by $\cos^2 \theta = 1 - \Delta \varepsilon_k / \varepsilon_{{\rm e}_b}$. This geometric constraint yields the fraction of forward momentum retained by a group $b$ electron after a channel $k$ collision:
\begin{equation}
\chi_{b,k} = \cos \theta = \sqrt{1 - \frac{\Delta \varepsilon_k}{\varepsilon_{{\rm e}_b}}}
\label{eqn:chicasc}
\end{equation}
While this equal-mass assumption is rigorous for electron-electron ionization encounters, excitation collisions target the entire atom. The resulting dipole scattering is more sharply forward-peaked, meaning Eq.\,\eqref{eqn:chicasc} somewhat overstates the de-beaming effect for excitation channels.

\begin{table*}[!ht]
  \center\fontsizetable
  \begin{threeparttable}
  \renewcommand{\arraystretch}{1.1}
    \caption{Electron multigroup chemical kinetics in argon. Each row denotes one reaction per destination group $j \in \{0, 1, \ldots, n_{\rm b}\}$; Reactions 3 to 6 additionally run over the parent group $b \in \{1, \ldots, n_{\rm b}\}$. }
    \label{tab:Ar_tab}
    \fontsizetable
    \begin{tabular*}{\textwidth}{l@{\extracolsep{\fill}}lll}
    \toprule
    No.~~~~& Reaction & Rate coefficient, m$^3$/s or 1/s~~\tnote{(a)}   & Source
\\
    \midrule
     1  & ${\rm e}_0^- + {\rm Ar} \rightarrow {\rm Ar}^+ + {\rm e}_0^- + {\rm e}_j^-$
        & $f_j \cdot k_1$         
        & Eqs.~\eqref{eqn:frunstep}--\eqref{eqn:jstar}, Table~\ref{tab:ei_splines_Ar} \\
     2  & ${\rm Ar}^+ + {\rm Ar} \rightarrow {\rm Ar}^+ + {\rm Ar}^+ + {\rm e}_j^-$
        & {\begin{minipage}[t]{6cm}\flushleft
        $ f_j \cdot \mu_{\rm Ar^+} |E| \cdot  \big(8.2\cdot 10^{-22} \cdot  \exp({-19900/(|E|/N)}) + 7.1 \cdot  10^{-21} \cdot  \exp({-46000/(|E|/N)})\big)$\end{minipage}}
        & Eqs.~\eqref{eqn:frunstep}--\eqref{eqn:jstar}, \eqref{eqn:muAr}, Ref.~\citenum{psst:1999:phelps} \\
     3  & ${\rm e}_b^- + {\rm Ar} \rightarrow {\rm e}_0^- + {\rm Ar}$
        & $f_0 \, \sigma_{\rm el}\bigl(\varepsilon_{{\rm e}_b}\bigr) V_{{\rm e}_b}$
        & Eq.~\eqref{eqn:scatrate}, Ref.~\citenum{lxcat:hayashi} \\
     4  & ${\rm e}_b^- + {\rm Ar} \rightarrow {\rm e}_j^- + {\rm Ar}^*_k$
        & $\big[ \chi_{b,k} \, g_j\bigl(\phi^{\rm birth}_{b,k}\bigr) + (1-\chi_{b,k}) f_j \big] \, \sigma_k\bigl(\varepsilon_{{\rm e}_b}\bigr) V_{{\rm e}_b}$
        & Eqs.~\eqref{eqn:cascade}, \eqref{eqn:scatrate_inel}, Ref.~\citenum{lxcat:hayashi} \\
     5  & ${\rm e}_b^- + {\rm Ar} \rightarrow {\rm Ar}^+ + {\rm e}_j^- + {\rm e}_{j'}^-$
        & $\big[ \chi_{b,{\rm iz}} \, g_j\bigl(\phi^{\rm birth}_{b,{\rm iz}}\bigr) + (1-\chi_{b,{\rm iz}}) f_j \big] f_{j'} \, \sigma_{\rm iz}\bigl(\varepsilon_{{\rm e}_b}\bigr) V_{{\rm e}_b}$\,\tnote{(b)}
        & Eqs.~\eqref{eqn:jstar}, \eqref{eqn:cascade}, \eqref{eqn:scatrate_inel}, Ref.~\citenum{lxcat:hayashi} \\
     6  & ${\rm e}_b^- \rightarrow {\rm e}_0^-$
        & $H\bigl(\varepsilon_{\rm born} - \varepsilon_{{\rm e}_b}\bigr) \cdot \max\bigl(0, -C_{\rm e} E V_{{\rm e}_b}\bigr) / \bigl(m_{\rm e} V_{{\rm e}_b}^2\bigr)$
        & Eq.~\eqref{eqn:turnrate} \\
     7  & ${\rm e}_0^- + {\rm Ar} \rightarrow {\rm e}_j^- + {\rm Ar}$
        & $\chi_0 \, g_j\bigl(\phi^{\rm birth}_0\bigr) \, f_{\rm run} \, |C_{\rm e}| / (m_{\rm e} \mu_{\rm e}^\star)$
        & Eqs.~\eqref{eqn:runrate}, \eqref{eqn:phibrun}, Table~\ref{tab:mueN} \\
    \bottomrule
    \end{tabular*}
\begin{tablenotes}
\item[{(a)}]  $\mu_{\rm Ar^+}$ is in m$^2$V$^{-1}$s$^{-1}$, $|E|/N$ in Townsends, and $|E|$ the absolute value of the electric field in V/m. In Reactions 3 to 7, all quantities are in SI units ($\sigma$ in m$^2$, $V_{{\rm e}_b}$ in m/s, $E$ in V/m, $C_{\rm e}$ in C, $m_{\rm e}$ in kg and $\mu_{\rm e}^\star$ in m$^{-1}$V$^{-1}$s$^{-1}$).   
\item[{(b)}]  Reaction 5 runs over both $j$ and $j'$ in $\{0, 1, \ldots, n_{\rm b}\}$: the bracket is the share of the primary sent to group $j$ and $f_{j'}$ the share of the secondary sent to group $j'$. Summed over $j'$, the rate coefficient gives the primary's deposit in group $j$; summed over both, one Ar$^+$ per ionizing collision.
\end{tablenotes}
   \end{threeparttable}
\end{table*}

This formulation allows the cascade of an electron from group $b$ to group $j$ to be expressed as a standard volumetric reaction. For an excitation channel $k$, a primary electron is consumed, a neutral is raised to the excited state, and a demoted primary electron is produced in group $j$:
\begin{equation}
{\rm e}_b^- + {\rm Ar} \; \rightarrow \; {\rm e}_j^- + {\rm Ar}^*_k
\label{eqn:cascreaction}
\end{equation}
For ionization channels, the primary electron is demoted to group $j$ just as in excitation, but with the added liberation of a secondary electron. The secondary is created at rest at the collision site and is therefore partitioned like any volumetrically liberated electron (Section~\ref{sec:runaway}): it joins group $j'$ with probability $f_{j'}$, the bulk group included:
\begin{equation}
{\rm e}_b^- + {\rm Ar} \; \rightarrow \; {\rm e}_j^- + {\rm e}_{j'}^- + {\rm Ar}^+
\label{eqn:cascreaction_iz}
\end{equation}
These processes correspond to Reactions 4 and 5 in Table \ref{tab:Ar_tab}. The cascade contribution to their macroscopic rate coefficient is:
\begin{equation}
k^{\rm casc}_{j \leftarrow b, k} = \chi_{b,k} \, g_j\bigl(\phi^{\rm birth}_{b,k}\bigr) \sigma_k\bigl(\varepsilon_{{\rm e}_b}\bigr) \, V_{{\rm e}_b}
\label{eqn:cascade}
\end{equation}
where the three factors represent the fraction of forward momentum surviving the collision, the destination share received by group $j$, and the collision rate coefficient $\sigma_k V_{{\rm e}_b}$, respectively. 

Notably, this cascade formulation conserves ballistic mass exactly. Because the destination weights sum to unity ($\sum_j g_j = 1$), summing Eq.\,\eqref{eqn:cascade} over all possible destination groups $j$ strictly recovers the total beam-conserving rate coefficient $\chi_{b,k} \sigma_k V_{{\rm e}_b} = \chi_{b,k} \nu^{\rm casc}_{b,k} / N_{\rm Ar}$. Thus, every ballistic electron consumed is fully accounted for downstream within the group family.

\subsection{Scattering sink}
\label{sec:scat}

Electrons are removed from their respective beams by two mechanisms: elastic collisions, and the fraction $1-\chi_{b,k}$ of inelastic collisions that deflect the electron too strongly for it to remain collimated (Sec.~\ref{sec:cascade}). The inelastic sink is therefore the exact complement of the degradation cascade. Together, these mechanisms define the de-beaming frequency of group $b$:
\begin{equation}
\nu^{\rm scat}_b = N_{\rm Ar}\, \sigma_{\rm deb}\bigl(\varepsilon_{{\rm e}_b}\bigr)\, V_{{\rm e}_b}
\label{eqn:nuscat}
\end{equation}
where the effective de-beaming cross-section $\sigma_{\rm deb}$ is given by:
\begin{equation}
\sigma_{\rm deb} = \sigma_{\rm el} + \sum_k \left( 1 - \chi_{b,k} \right) \sigma_k
\label{eqn:sigmadeb}
\end{equation}
Each term in $\sigma_{\rm deb}$ counts collisions in proportion to the forward momentum they destroy. The inelastic terms require the weight $1-\chi_{b,k}$ to capture this loss. The elastic term requires no additional weight because $\sigma_{\rm el}$ is the momentum-transfer cross-section, $\int(1-\cos\theta)\,{\rm d}\sigma$, with ${\rm d}\sigma$ the differential elastic cross-section for scattering through the angle $\theta$; this weighting discounts grazing deflections and doubles the contribution of backscattering events. A group thus loses its beam over a de-beaming length of $1/(N_{\rm Ar}\sigma_{\rm deb})$.

Upon scattering, the electron is treated as if it were newly liberated at the collision site and is reassigned according to the runaway partition. For an elastic collision, this reads:
\begin{equation}
{\rm e}_b^- + {\rm Ar} \; \rightarrow \; {\rm e}_j^- + {\rm Ar}
\label{eqn:scatreaction}
\end{equation}
which appears in Table~\ref{tab:Ar_tab} as Reaction 3, the share $f_{\rm run}$ returning to group $b$ being a null operation, as explained below.

For elastic collisions, the scattering rate coefficient is:
\begin{equation}
k^{\rm scat}_{j \leftarrow b} = \delta_{j 0} \, f_0 \, \sigma_{\rm el}\bigl(\varepsilon_{{\rm e}_b}\bigr) \, V_{{\rm e}_b}
\label{eqn:scatrate}
\end{equation}
where $\delta_{j0}$ is the Kronecker delta, unity for the bulk group and zero otherwise. Because an elastic collision removes at most a tiny fraction of the electron's energy ($2m_{\rm e}/m_{\rm Ar} \approx 3\times10^{-5}$), it never contributes to the degradation cascade, and the share $f_{\rm run}$ that runs away again returns to group $b$ unchanged. This null operation is why Eq.\,\eqref{eqn:scatrate} includes only the bulk fraction $f_0$, and why the scattering rates summed over all destinations fall short of $\sigma_{\rm deb} V_{{\rm e}_b}$ by the share $f_{\rm run}$ of its elastic part. Returning these electrons to group $b$ slightly overestimates their directed momentum, whereas using an $f_{\rm run}$ derived for electrons starting from rest slightly underestimates the runaway probability of a fast, deflected electron. The two errors partly offset.

For an inelastic collision, the products are instead those of Eqs.\,\eqref{eqn:cascreaction} and \eqref{eqn:cascreaction_iz}, and the scattering rate coefficient is:
\begin{equation}
k^{\rm scat}_{j \leftarrow b, k} = f_j \left( 1 - \chi_{b,k} \right) \sigma_k\bigl(\varepsilon_{{\rm e}_b}\bigr) \, V_{{\rm e}_b}
\label{eqn:scatrate_inel}
\end{equation}
which appears in Table~\ref{tab:Ar_tab} as the term $(1-\chi_{b,k}) f_j$ of Reactions 4 and 5. Having lost an energy $\Delta\varepsilon_k$, the inelastically scattered electron can no longer return to group $b$. Its residual kinetic energy, $\varepsilon_{{\rm e}_b}-\Delta\varepsilon_k$, lacks a preferred direction. According to Eq.\,\eqref{eqn:phibirth}, the electron therefore enters the partition with zero forward energy, at a birth potential $\phi_{\rm birth}=\phi$. The share $f_{\rm run}$ is then re-collimated by the field, while the share $f_0$ joins the thermal bulk. Because a unidirectional beam cannot store this undirected residual energy, it is credited entirely to the bulk energy equation of Section~\ref{sec:bulk}, regardless of how the mass is partitioned. Since the field cannot convert transverse energy into forward motion between collisions, this choice affects only where the residual energy is stored, not how the re-collimated share regains energy from the field.

\subsection{Turning point and return to the bulk group}
\label{sec:turn}

A group entering a region of falling potential decelerates correctly; what it cannot represent is the end of that deceleration. Where the potential falls back to the group's own birth potential, its members have given back everything the field lent them and reverse, and since every group travels along $+x$, the mass must leave there.

No collision carries the transfer, so the reaction, Reaction 6 of Table~\ref{tab:Ar_tab}, is unimolecular:
\begin{equation}
{\rm e}_b^- \; \rightarrow \; {\rm e}_0^-
\label{eqn:turnreaction}
\end{equation}
with the bulk group as its sole destination.

The rate of this transfer follows from the deceleration itself. Between collisions, an electron of group $b$ obeys $m_{\rm e} V_{{\rm e}_b}\, \partial V_{{\rm e}_b}/\partial x = C_{\rm e} E$, so that where the field decelerates the group, that is where $C_{\rm e} E V_{{\rm e}_b} < 0$, its speed falls at the fractional rate $-C_{\rm e} E/\bigl(m_{\rm e} V_{{\rm e}_b}\bigr)$. We take this rate as the rate coefficient once the energy of the group has fallen below the birth energy $\varepsilon_{\rm born} = \tfrac{3}{2} k_{\rm B} T_{\rm born}$:
\begin{equation}
k^{\rm turn}_b = H\!\left( \varepsilon_{\rm born} - \varepsilon_{{\rm e}_b} \right) \frac{\max\left( 0, \, -C_{\rm e} E V_{{\rm e}_b} \right)}{m_{\rm e} V_{{\rm e}_b}^2}
\label{eqn:turnrate}
\end{equation}
where $H$ is the Heaviside step function. The numerator, where positive, is the power the field extracts from an electron of the group, and the denominator is twice its kinetic energy. This rate coefficient removes exactly the compression of the decelerating beam: in steady state, Eq.\,\eqref{eqn:masstransport_ej} with this sink alone gives $\partial\bigl(\rho_{{\rm e}_b} V_{{\rm e}_b}\bigr)/\partial x = -k^{\rm turn}_b \rho_{{\rm e}_b} = \rho_{{\rm e}_b}\, \partial V_{{\rm e}_b}/\partial x$, and hence $\partial \rho_{{\rm e}_b}/\partial x = 0$. Below $\varepsilon_{\rm born}$ the density of the group therefore stays constant while its flux falls in proportion to $V_{{\rm e}_b}$, reaching zero exactly at the turning point, where the rate coefficient diverges. The whole of the group is thus handed to the bulk over the last $\varepsilon_{\rm born}$ of its deceleration, and no adjustable constant is involved. The rate coefficient vanishes wherever the field accelerates the group, so that electrons that have just joined a group with little energy are not returned to the bulk.

\begin{table}[!t]
  \center\fontsizetable
  \begin{threeparttable}
    \caption{Coordinates of spline control points for argon electron impact reaction {${\rm e}_0^- + {\rm Ar} \rightarrow {\rm Ar}^+ + {\rm e}_0^- + {\rm e}_j^-$}.\tnote{(a)}}
    \label{tab:ei_splines_Ar} 
    \fontsizetable
    \begin{tabular*}{\columnwidth}{@{\extracolsep{\fill}}ll@{}}
    \toprule
     ${\rm ln}~T_{\rm e}$ & ${\rm ln}~k_1$ \\
    \midrule
      \begin{minipage}[t]{0.47\columnwidth}\raggedright  
      10.42099
      10.56124
      10.59131
      10.65607
      10.70699
      10.7947
      10.92818
      11.12563
      11.45201
      13.11916
      16.1181
    \end{minipage}  & \begin{minipage}[t]{0.46\columnwidth}\raggedright 
      -76.65280
      -62.60476
      -58.59835
      -48.07229
      -43.27593
      -39.10465
      -36.06700
      -34.02993
      -32.4686
      -29.9596
      -29.25725
    \end{minipage} \alb         
    \bottomrule
    \end{tabular*}
    \begin{tablenotes}
    \item[{(a)}] The rate coefficient $k_1$ has units of $\textrm{m}^3\cdot \textrm{s}^{-1}$.
The electron temperature $T_{\rm e}$ is in Kelvin. The discrete control points were generated by evaluating cross-sections from the Hayashi database using BOLSIG- and fitting the resulting temperature-dependent rate data using a monotone cubic spline with a relative error not exceeding 3\%. 
\end{tablenotes}
   \end{threeparttable}
\end{table}

\subsection{Thermal runaway and return to the ballistic groups}
\label{sec:thermalrunaway}

Without a mechanism to return bulk electrons to the ballistic family, electrons in regions where the electric field exceeds $E_{\rm crit}$ could artificially accumulate in the bulk. This violates the partition rule of Eq.\,\eqref{eqn:frunstep}, which dictates that no sustained thermal population can exist in such a regime.

This unphysical pileup is resolved by the reverse of Reaction 3: thermal runaway. A collision re-randomizes a bulk electron's trajectory; if scattered favorably, the electric field may accelerate it past the peak of the friction curve before its next collision. This two-body process, Reaction 7 of Table~\ref{tab:Ar_tab}, leaves the neutral target unchanged:
\begin{equation}
{\rm e}_0^- + {\rm Ar} \; \rightarrow \; {\rm e}_j^- + {\rm Ar}
\label{eqn:runreaction}
\end{equation}
The rate coefficient is the product of an attempt frequency and a success probability. The attempt frequency is the momentum-transfer collision frequency of the bulk electrons, obtained from the reduced mobility $\mu_{\rm e}^\star = \mu_{\rm e} N$, with $\mu_{\rm e}$ the bulk electron mobility and $N$ the total gas number density, as $|C_{\rm e}| N_{\rm Ar}/(m_{\rm e} \mu_{\rm e}^\star)$ \cite{pf:2024:parent}. Being averaged over the electron distribution and including the momentum lost in inelastic collisions, this frequency exceeds the elastic rate $\sigma_{\rm el}(\overline{\varepsilon_{\rm e}}) \overline{q_{\rm e}} N_{\rm Ar}$, with $\overline{q_{\rm e}}$ the mean thermal speed of the bulk electrons (Eq.\,\eqref{eqn:qbar}), by a factor of 3 to 6 at bulk temperatures between 30 and 170~eV, where runaway is active; every such collision randomizes the direction of the electron and presents an attempt. The success probability is the product of two factors. The first is $f_{\rm run}$: a low-energy thermal electron must climb the entire friction curve, just like a newly liberated electron, which makes $f_{\rm run}$ the appropriate partition weight. The second accounts for direction. Only an electron scattered into the accelerating sense can be re-collimated promptly, and elastic scattering from argon at a few electron-volts is close to isotropic, so half the post-collision directions qualify. This is written as a user-defined constant $\chi_0$, set to $0.5$ here, which plays for a thermal electron the role that $\chi_{b,k}$ plays for a ballistic one. The runaway rate coefficient into destination group $j$ is therefore:
\begin{equation}
k^{\rm run}_{j \leftarrow 0} = \chi_0 \, g_j\bigl(\phi^{\rm birth}_0\bigr) f_{\rm run} \frac{|C_{\rm e}|}{m_{\rm e} \mu_{\rm e}^\star}
\label{eqn:runrate}
\end{equation}
These runaway electrons do not emerge with zero forward energy. Because the field selectively extracts electrons whose post-collision trajectories align parallel to the accelerating force, the successful runaway population is biased toward forward motion. We approximate this initial forward energy as their pre-collision thermal energy ($\varepsilon_{+x} \approx \overline{\varepsilon_{\rm e}}$). Substituting this into Eq.\,\eqref{eqn:phibirth} yields their effective birth potential:
\begin{equation}
\phi^{\rm birth}_0 = \phi - \frac{\tfrac{3}{2} k_{\rm B} T_{\rm e}}{|C_{\rm e}|}
\label{eqn:phibrun}
\end{equation}
Below $E_{\rm crit}$, $f_{\rm run}$ vanishes, rendering this reaction inert and leaving the thermal population of the negative glow undisturbed. Above $E_{\rm crit}$, the runaway fraction approaches unity, the reaction rate converges to the fraction $\chi_0$ of the momentum-transfer collision frequency, and the bulk population is rapidly swept back into the ballistic groups.

\subsection{Ballistic source term}
\label{sec:ballisticsource}

Collecting the mechanisms above, Table~\ref{tab:Ar_tab} evaluated for a ballistic group gives the source of Eq.\,\eqref{eqn:masstransport_ej}:
\begin{align}
W_{{\rm e}_j} &= m_{\rm e} f_j \Bigg[ k_1 N_{{\rm e}_0} N_{\rm Ar} + k_2 N_{\rm Ar^+} N_{\rm Ar}
 + N_{\rm Ar} \sum_{b=1}^{n_{\rm b}} N_{{\rm e}_b} V_{{\rm e}_b} \, \sigma_{\rm iz}\bigl(\varepsilon_{{\rm e}_b}\bigr) \Bigg] \nonumber \\
 &\quad + m_{\rm e} N_{\rm Ar} \sum_{b=1}^{n_{\rm b}} N_{{\rm e}_b} V_{{\rm e}_b}
 \sum_k \Big[ \chi_{b,k} \, g_j\bigl(\phi^{\rm birth}_{b,k}\bigr) + \left( 1 - \chi_{b,k} \right) f_j \Big] \sigma_k\bigl(\varepsilon_{{\rm e}_b}\bigr) \nonumber \\
 &\quad - m_{\rm e} N_{\rm Ar} N_{{\rm e}_j} V_{{\rm e}_j} \Bigg[ f_0 \, \sigma_{\rm el}\bigl(\varepsilon_{{\rm e}_j}\bigr) + \sum_k \sigma_k\bigl(\varepsilon_{{\rm e}_j}\bigr) \Bigg]
 \; - \; \rho_{{\rm e}_j} k^{\rm turn}_j \nonumber \\
 &\quad + \chi_0 \, m_{\rm e} N_{\rm Ar} N_{{\rm e}_0} \, f_{\rm run} \frac{|C_{\rm e}|}{m_{\rm e} \mu_{\rm e}^\star} \; g_j\bigl(\phi^{\rm birth}_0\bigr)
 \quad \text{for } j \in \{1, \ldots, n_{\rm b}\}
 \label{eqn:Wej}
\end{align}
Here $N_{{\rm e}_j} = \rho_{{\rm e}_j}/m_{\rm e}$ is the number density of group $j$, the bulk group included, $N_{\rm Ar^+}$ is the number density of the argon ions, and $k_1$ and $k_2$ are the electron- and ion-impact ionization rate coefficients of Reactions 1 and 2, the former a function of the bulk temperature $T_{\rm e}$ (Section~\ref{sec:bulk}). The first line is the share $f_j$ of the electrons liberated at rest; the second, the ballistic electrons received from every group $b$ through inelastic collisions, demoted by the cascade (share $\chi_{b,k}$, at the birth potential of Eq.\,\eqref{eqn:phibcasc}) or deflected and re-partitioned (share $1-\chi_{b,k}$); the third, the losses of the group to collisions and to the turning point; and the fourth, the gain from thermal runaway, at the birth potential of Eq.\,\eqref{eqn:phibrun}. The share $f_{\rm run}$ of an elastically de-beamed electron returns to the group it left with its energy unchanged, a null operation, which is why only $f_0 \sigma_{\rm el}$ appears in the loss bracket. The group $j = n_{\rm b}$ is fed only by cathode emission through the boundary condition of Section \ref{sec:bcs}, so $f_{n_{\rm b}} = g_{n_{\rm b}} = 0$ and its source contains the loss terms alone.

\section{Bulk Electron Group}
\label{sec:bulk}

Since only the bulk group sustains random thermal motion, its thermal properties carry the subscript ${\rm e}$ without a group index, as in the temperature $T_{\rm e}$. The bulk electrons are governed by the mass continuity equation:
\begin{equation}
\frac{\partial \rho_{{\rm e}_0}}{\partial t} + \frac{\partial }{\partial x} (\rho_{{\rm e}_0} V_{{\rm e}_0}) = W_{{\rm e}_0}
\label{eqn:masstransport_e1}
\end{equation}
where $\rho_{{\rm e}_0}$ is the bulk mass density, $V_{{\rm e}_0}$ is the bulk macroscopic velocity, and $W_{{\rm e}_0}$ is the net chemical source term. 

This source term accumulates all creation and destruction processes relevant to the bulk electrons, corresponding to specific rows in the reaction mechanism detailed in Table \ref{tab:Ar_tab}. Evaluating these contributions yields:
\begin{align}
W_{{\rm e}_0} &= m_{\rm e} f_0 \Bigg[ k_1 N_{{\rm e}_0} N_{\rm Ar} + k_2 N_{\rm Ar^+} N_{\rm Ar}
 + N_{\rm Ar} \sum_{b=1}^{n_{\rm b}} N_{{\rm e}_b} V_{{\rm e}_b} \, \sigma_{\rm iz}\bigl(\varepsilon_{{\rm e}_b}\bigr) \Bigg] \nonumber \\
 &\quad + m_{\rm e} f_0 N_{\rm Ar} \sum_{b=1}^{n_{\rm b}} N_{{\rm e}_b} V_{{\rm e}_b}
 \Bigg[ \sigma_{\rm el}\bigl(\varepsilon_{{\rm e}_b}\bigr) + \sum_k \left( 1 - \chi_{b,k} \right) \sigma_k\bigl(\varepsilon_{{\rm e}_b}\bigr) \Bigg] \nonumber \\
 &\quad + \sum_{b=1}^{n_{\rm b}} \rho_{{\rm e}_b} k^{\rm turn}_b
 \; - \; \chi_0 \, m_{\rm e} N_{\rm Ar} N_{{\rm e}_0} \, f_{\rm run} \frac{|C_{\rm e}|}{m_{\rm e} \mu_{\rm e}^\star}
 \label{eqn:We0}
\end{align}
where the index $b$ runs over the parent ballistic groups and the index $k$ over all inelastic channels, including ionization. The cross-sections $\sigma_{\rm iz}$, $\sigma_{\rm el}$ and $\sigma_k$ are those of electron-impact ionization, of elastic momentum transfer, and of inelastic channel $k$, and they are evaluated at the kinetic energy $\varepsilon_{{\rm e}_b}$ of group $b$, given by Eq.~\eqref{eqn:eps_group}. The first line collects electrons liberated at rest via Reactions 1, 2, and 5, of which the bulk group receives the fraction $f_0$. The second line represents the fraction $f_0$ of ballistic electrons de-beamed by elastic and strongly deflecting inelastic collisions (Reactions 3 to 5). The third line accounts for mass gained at kinematic turning points (Reaction 6) and mass lost to thermal runaway (Reaction 7). The partition fractions $f_0$ and $f_{\rm run}$, the retained momentum fraction $\chi_{b,k}$, the re-collimation fraction $\chi_0$, and the turning coefficient $k^{\rm turn}_b$ are those of Section~\ref{sec:ballistic}. Because the primary ionizing bulk electron in Reaction 1 survives the collision, the bulk group suffers no net collisional mass loss from this process; its contribution enters purely through the electron-impact ionization rate coefficient $k_1$. This rate coefficient is a function of the bulk electron temperature $T_{\rm e}$, obtained using BOLSIG- and parameterized in Table \ref{tab:ei_splines_Ar}. The coefficient $k_2$ is the corresponding ion-impact rate coefficient from Reaction 2.

Adding Eq.\,\eqref{eqn:We0} to the sum of Eq.\,\eqref{eqn:Wej} over the ballistic groups, the transfer, turning-point and runaway terms cancel identically and only the liberation terms remain, so electrons are created only by ionization.

Neglecting electron inertia but retaining the pressure gradient, the bulk electron velocity is expressed as:
\begin{equation}
V_{{\rm e}_0} = - \mu_{\rm e} E - \frac{\mu_{\rm e}}{|C_{\rm e}|
N_{{\rm e}_0}} \frac{\partial P_{\rm e}}{\partial x}
\label{eqn:Vk_e1}
\end{equation}
where $\mu_{\rm e}$ is the electron mobility (tabulated in Table \ref{tab:mueN}), and $P_{\rm e} = N_{{\rm e}_0} k_{\rm B} T_{\rm e}$ is the partial pressure. The local bulk electron temperature evolves according to the energy equation:
\begin{align}
\frac{\partial }{\partial t} (\rho_{{\rm e}_0} e_{\rm e}) + \frac{\partial }{\partial x} \left( \rho_{{\rm e}_0} V_{{\rm e}_0} h_{\rm e}  - \kappa_{\rm e} \frac{\partial T_{\rm e}}{\partial x}  \right) =& Q_{{\rm e}_0}^{\rm born} + Q_{{\rm e}_0}^{\rm beam} - Q_{{\rm e}_0}^{\rm run} \nonumber \\ 
 &+  C_{\rm e} N_{{\rm e}_0} E V_{{\rm e}_0} - Q_{{\rm e}_0}^{\rm relax}
\label{eqn:electronenergy_e1}
\end{align}
where $e_{\rm e}(T_{\rm e}) = \frac{3}{2} \frac{k_{\rm B}}{m_{\rm e}} T_{\rm e}$, $h_{\rm e} = \frac{5}{2} \frac{k_{\rm B}}{m_{\rm e}} T_{\rm e}$, and the thermal conductivity is $\kappa_{\rm e} = \frac{5}{2} N_{{\rm e}_0} k_{\rm B}^2 T_{\rm e} \mu_{\rm e} / |C_{\rm e}|$. The source terms on the right-hand side represent Joule heating ($C_{\rm e} N_{{\rm e}_0} E V_{{\rm e}_0}$), net collisional energy transfer with heavy particles ($Q_{{\rm e}_0}^{\rm relax}$), and the energy carried by electrons entering the bulk at creation ($Q_{{\rm e}_0}^{\rm born}$) or from the ballistic groups ($Q_{{\rm e}_0}^{\rm beam}$), or leaving it by runaway ($Q_{{\rm e}_0}^{\rm run}$).

\begin{table}[!t]
  \center\fontsizetable
  \begin{threeparttable}
    \caption{Spline control points for the reduced electron mobility $\mu_{\rm e}^\star=\mu_{\rm e} N$ in argon.\tnote{(a)}}
    \label{tab:mueN}
    \begin{tabular*}{\columnwidth}{@{\extracolsep{\fill}}ll@{}}
    \toprule
    $\ln(T_{\rm e})$ & $\ln(\mu_{\rm e} N)$ \\ 
    \midrule
    \begin{minipage}[t]{0.45\columnwidth}\raggedright
      5.714736
      6.414312
      7.157508
      8.591843
      9.684635
      10.53106
      10.66237
      10.84248
      12.0376
      13.77185
      16.1181
    \end{minipage} & \begin{minipage}[t]{0.46\columnwidth}\raggedright
      60.06979
      60.4004
      60.41304
      58.21029
      56.57183
      55.20634
      55.16945
      54.98991
      54.67693
      54.32447
      54.2868
    \end{minipage} 
    \alb
    \bottomrule
    \end{tabular*}
    \begin{tablenotes}
    \item[{(a)}] 
$T_{\rm e}$ has units of $\text{K}$, and  $\mu_{\rm e} N$ has units of $\text{m}^{-1}\text{V}^{-1}\text{s}^{-1}$.
Such were computed using BOLSIG- and the Hayashi database (LXCat) and fitted within 3\% accuracy using monotone spline interpolation. 
\end{tablenotes}
   \end{threeparttable}
\end{table}

The term $Q_{{\rm e}_0}^{\rm born}$ represents the thermal energy deposited into the bulk by electrons liberated at rest, credited at an assumed creation energy $e_{\rm e}(T_{\rm born})$:
\begin{align}
Q_{{\rm e}_0}^{\rm born} = &m_{\rm e} f_0 \, e_{\rm e}(T_{\rm born}) \Bigg[ k_1 N_{{\rm e}_0} N_{\rm Ar} + k_2 N_{\rm Ar^+} N_{\rm Ar}\nonumber \\
&+ N_{\rm Ar} \sum_{b=1}^{n_{\rm b}} N_{{\rm e}_b} V_{{\rm e}_b} \, \sigma_{\rm iz}\bigl(\varepsilon_{{\rm e}_b}\bigr) \Bigg]
\label{eqn:Qborn}
\end{align}
The bracketed expression corresponds to the first line of Eq.\,\eqref{eqn:We0}.

Conversely, $Q_{{\rm e}_0}^{\rm beam}$ captures the kinetic energy thermalized into the bulk when energetic electrons are scattered out of the ballistic groups. Because these electrons carry significant energy prior to scattering, the heat addition is:
\begin{align}
Q_{{\rm e}_0}^{\rm beam} = &\sum_{b=1}^{n_{\rm b}} \rho_{{\rm e}_b} N_{\rm Ar} V_{{\rm e}_b} \left[ f_0 \, \sigma_{\rm el} \frac{\varepsilon_{{\rm e}_b}}{m_{\rm e}} + \sum_k \left( 1 - \chi_{b,k} \right) \sigma_k \frac{\varepsilon_{{\rm e}_b} - \Delta \varepsilon_k}{m_{\rm e}} \right] \nonumber \\
&+ \sum_{b=1}^{n_{\rm b}} \rho_{{\rm e}_b} \, k^{\rm turn}_b \, \frac{\varepsilon_{{\rm e}_b}}{m_{\rm e}}
\label{eqn:Qbeam}
\end{align}
The specific mechanisms governing these transfers are detailed in Section \ref{sec:ballistic}.

Bulk electrons also lose energy when they undergo thermal runaway (Reaction 7) and join a ballistic group. The energy removed by this escaping mass is:
\begin{equation}
Q_{{\rm e}_0}^{\rm run} = \chi_0 \, \rho_{{\rm e}_0} N_{\rm Ar} \, f_{\rm run} \frac{|C_{\rm e}|}{m_{\rm e} \mu_{\rm e}^\star} \, e_{\rm e}(T_{\rm e})
\label{eqn:Qrun}
\end{equation}
where the departing electrons are assumed to carry away exactly the mean thermal specific energy they held prior to runaway.

Following Ref.~\citenum{pf:2024:parent}, the net inelastic cooling and superelastic heating rates are parameterized entirely through the species-dependent reduced electric field $E^\star$ and reduced electron mobility $\mu_{\rm e}^\star$. This formulation circumvents the need to track individual cross-sections for every transition. For a pure argon background, the aggregate collisional relaxation term is:
\begin{align}
Q_{{\rm e}_0}^{\rm relax} &=  |C_{\rm e}|
N_{{\rm e}_0} N_{\rm Ar} \mu_{\rm e}^\star \left( \bigl[E^\star(T_{\rm e})\bigr]^2 -\frac{3 k_{\rm B} (T_{\rm e}-T_{\rm ref})}{ m_{\rm Ar} \bigl(\mu_{\rm e}^\star\bigr)^2} \right) \nonumber \\
&\quad +  \frac{3  k_{\rm B} |C_{\rm e}|
N_{{\rm e}_0} N_{\rm Ar} (T_{\rm e}-T)}{ m_{\rm Ar} \mu_{\rm e}^\star} \nonumber \\
&\quad +  N_{{\rm e}_0} N_{\rm Ar^+} (T_{\rm e}-T ) \frac{6 k_{\rm B} C_{\rm e}^2 C_{\rm Ar^+}^2 \ln \Lambda}{ \pi^3 \epsilon_0^2 m_{\rm e} m_{\rm Ar^+} \overline{q_{\rm e}}^3}
\label{eqn:Qrelax}
\end{align}
On the right-hand side, the first term captures the net inelastic transfer with neutrals, the second describes elastic transfer with neutrals, and the third dictates energy exchange with ions via Coulomb collisions. Here, $m_{\rm Ar}$ and $m_{\rm Ar^+}$ are the masses of the argon atom and ion, $C_{\rm Ar^+}$ is the charge of the ion, $\epsilon_0$ is the permittivity of free space, $T$ is the gas temperature, $T_{\rm ref} = 300$~K is the reference temperature of the elastic relaxation term, and $\overline{q_{\rm e}}$ is the mean thermal electron speed:
\begin{equation}
\overline{q_{\rm e}} = \sqrt{\frac{8 k_{\rm B} T_{\rm e}}{\pi m_{\rm e}}}
\label{eqn:qbar}
\end{equation}
The Coulomb logarithm follows the NRL Formulary approximation \cite{nrl:2002:huba}:
\begin{equation}
\ln \Lambda = 23 - \ln \left( N_{{\rm e}_0}^{0.5} T_{\rm e}^{-1.5} \right)
\end{equation}
with $T_{\rm e}$ given in electron-volts and $N_{{\rm e}_0}$ in $\text{cm}^{-3}$. 

Finally, the effective reduced electric field $E^\star$ and reduced electron mobility $\mu_{\rm e}^\star$ are computed from the local bulk electron temperature. These transport parameters are reconstructed via monotone spline interpolation using control points derived from the Hayashi database (LXCat) via BOLSIG-, as tabulated in Tables \ref{tab:Estar} and \ref{tab:mueN}. This closure assumes a local distribution for the bulk electrons, an assumption known to fail when a single fluid represents the whole population in the negative glow \cite{psst:2024:zhou}. Here the energetic electrons, and the ionization they produce, are carried by the ballistic groups, so the bulk no longer needs an inflated temperature to sustain the discharge and the local closure is correspondingly more defensible, though still approximate.

\begin{table}[!t]
  \center\fontsizetable
  \begin{threeparttable}
    \caption{Spline control points for the reduced electric field $E^\star$ in argon.\tnote{(a)}}
    \label{tab:Estar} 
    \begin{tabular*}{\columnwidth}{@{\extracolsep{\fill}}ll@{}}
    \toprule
    $\ln(T_{\rm e})$ & $\ln(E^\star)$ \\ 
    \midrule
    \begin{minipage}[t]{0.43\columnwidth}\raggedright
      5.714736
      5.728932
      5.863773
      5.922798
      7.774939
      7.784937
      10.23766
      10.54873
      10.59276
      10.72383
      10.93272
      11.12947
      12.44404
      16.1181
    \end{minipage} & \begin{minipage}[t]{0.46\columnwidth}\raggedright
      -56.44688
      -56.08551
      -55.26188
      -55.13269
      -52.97587
      -52.94037
      -47.42898
      -46.66301
      -46.43841
      -44.93257
      -43.32603
      -42.51776
      -40.5886
      -38.16674
    \end{minipage} 
    \alb
    \bottomrule
    \end{tabular*}
    \begin{tablenotes}
    \item[{(a)}] $T_{\rm e}$ has units of $\text{K}$,  $E^\star$ has units of $\text{V}\cdot\text{m}^2$.
Such were obtained with BOLSIG- using the Hayashi database (LXCat) and parameterized via monotone cubic spline interpolation within 3\% accuracy. 
\end{tablenotes}
   \end{threeparttable}
\end{table}

\section{Heavy Particle Transport Equations}
\label{sec:heavy}

The mass conservation equation for the argon ions is given by:
\begin{equation}
\frac{\partial \rho_{\rm Ar^+}}{\partial t} + \frac{\partial }{\partial x} (\rho_{\rm Ar^+} V_{\rm Ar^+}) = W_{\rm Ar^+}
\end{equation}
where $\rho_{\rm Ar^+}$ is the ion mass density, $V_{\rm Ar^+}$ is the macroscopic ion velocity along the $x$-direction, and $W_{\rm Ar^+}$ is the net mass generation rate from the volumetric reactions detailed in Table~\ref{tab:Ar_tab}. This source term explicitly incorporates ``hot'' ion impact ionization (Reaction 2), utilizing a spatial coefficient parameterized from Appendix B of Ref.~\citenum{psst:1999:phelps}.

The macroscopic velocity $V_{\rm Ar^+}$ captures both drift and diffusion through a background gas at rest:
\begin{equation}
V_{\rm Ar^+} = \mu_{\rm Ar^+} E - \frac{\mu_{\rm Ar^+}}{C_{\rm Ar^+} N_{\rm Ar^+}} \frac{\partial P_{\rm Ar^+}}{\partial x}
\end{equation}
where $\mu_{\rm Ar^+}$ is the ion mobility and $C_{\rm Ar^+}$ is the elementary ion charge. The ion partial pressure is defined as $P_{\rm Ar^+} = N_{\rm Ar^+} k_{\rm B} T$, with $N_{\rm Ar^+}$ being the ion number density and $T$ assuming thermal equilibrium with the background gas.

Assuming a not too elevated background gas temperature, the baseline zero-field mobility for $\text{Ar}^+$ in Ar can be taken as inversely proportional to the total gas number density $N$. To account for symmetric charge-exchange collisions at elevated electric fields, this mobility is corrected with an expression that closely approximates the Ar$^+$ drift velocity formula of Phelps and Petrovi\'c (Eq.~(B9) of Ref.~\citenum{psst:1999:phelps}), agreeing with it to within 4\% for $|E|/N$ above 1000~Td. The corrected mobility $\mu_{\rm Ar^+}$ (in $\text{m}^2 \text{V}^{-1} \text{s}^{-1}$) takes the form:
\begin{equation}
\mu_{\rm Ar^+} = N^{-1} \min\left(  4.08 \times 10^{21},~  0.40\cdot m_{\rm Ar}^{-0.5} \cdot (|E|/N)^{-0.5} \right)
\label{eqn:muAr}
\end{equation}
where $N$ is the total gas number density ($\text{m}^{-3}$), $|E|/N$ is the reduced electric field ($\text{V}\cdot\text{m}^2$), and $m_{\rm Ar}$ is the argon atomic mass (kg).

\section{Boundary Conditions}
\label{sec:bcs}

We assume no surface catalysis for the neutrals, while electrons and ions are assumed to recombine fully at the boundaries. Let $\eta$ denote the spatial coordinate normal to the surface, pointing positively toward the plasma interior. Under this convention, a surface where the normal electric field component $E_\eta < 0$ acts as a cathode, whereas $E_\eta > 0$ defines an anode. 

Secondary electron emission is governed by a user-defined coefficient $\gamma_{\rm e}$, set for each test case in Section~\ref{sec:tests}. Because all secondary electrons are exclusively assigned to the emission group $n_{\rm b}$, their flux at the cathode is explicitly prescribed:
\begin{equation}
 \rho_{{\rm e}_{n_{\rm b}}} V_{{\rm e}_{n_{\rm b}}}  =  m_{\rm e} \gamma_{\rm e} \, \mu_{\rm Ar^+} |E_\eta| \, N_{\rm Ar^+}  \quad \text{for } E_\eta < 0
\label{eqn:emitbc}
\end{equation}
Imposing this condition directly on the flux ensures that exactly $\gamma_{\rm e}$ electrons are emitted per incident ion, independent of the numerical grid resolution. 

The mean group potential of the secondary group is initialized based on the emission kinematics. Secondaries are generated at the local surface potential $\phi_{\rm cathode}$ with a uniform emission energy $\varepsilon_{\rm emit}$. According to Eq.\,\eqref{eqn:phibirth}, their corresponding birth potential is $\phi_{\rm emit} = \phi_{\rm cathode} - \varepsilon_{\rm emit}/|C_{\rm e}|$, which sets its mean group potential $\overline{\phi}_{n_{\rm b}} = \phi_{\rm emit}$. Because group $n_{\rm b}$ is strictly reserved for surface emission and receives no volumetric mass, the trajectory, energy, and velocity of these secondary electrons through the sheath and plasma follow Newton's laws exactly.

Conversely, the bulk electron group receives no mass from the cathode. Physically, the cathode emits directly into a high-field sheath where a thermalized population cannot be sustained; thus, the bulk density is forced to zero:
\begin{equation}
N_{{\rm e}_0} = 0 \quad \text{for } E_\eta < 0
\label{eqn:emitbc_e0}
\end{equation}
At an anode ($E_\eta > 0$), the bulk electron group is instead assigned a zero-gradient Neumann condition:
\begin{equation}
\frac{\partial N_{{\rm e}_0}}{\partial \eta} = 0 \quad \text{for } E_\eta > 0
\label{eqn:zerograd_e0}
\end{equation}
The ballistic groups require no boundary condition at the anode, as their hyperbolic advection is strictly one-sided (either outflowing or null). 

For the ions, these boundary conditions are reversed. A zero-gradient condition is applied at the cathode:
\begin{equation}
\frac{\partial N_{\rm Ar^+}}{\partial \eta} 
= 0 \quad \text{for } E_\eta < 0
\end{equation}
and a homogeneous Dirichlet condition is enforced at the anode:
\begin{equation}
N_{\rm Ar^+} = 0 \quad \text{for } E_\eta > 0
\end{equation}
The bulk electron temperature is subjected to a zero-gradient condition at all boundaries:
\begin{equation}
\frac{\partial T_{\rm e}}{\partial \eta} = 0
\label{eqn:zerograd_Te0}
\end{equation}
Finally, to ensure numerical stability, the sign of $E_\eta$ used to switch these boundary conditions is evaluated a few nodes into the plasma interior rather than strictly at the physical face.

\subsection{Absorbing Anode Wall Condition}
\label{sec:nonreflecting}

We also formulate an alternative electron density boundary condition specifically for a perfectly absorbing anode. At such a wall, the escaping one-sided flux follows a half-Maxwellian distribution. In a steady state, this kinetic loss must balance the mass delivered by diffusion across the final half-cell of the domain:
\begin{equation}
\frac{1}{4} N_{{\rm e}_0} \, \overline{q_{\rm e}} = D_{\rm e} \, \frac{N_{{\rm e}_0}^{\rm nw} - N_{{\rm e}_0}}{\Delta x / 2} \qquad \text{for } E_\eta > 0
\label{eqn:nonrefbalance}
\end{equation}
where $\Delta x$ is the cell width, $N_{{\rm e}_0}$ and $N_{{\rm e}_0}^{\rm nw}$ are the electron densities at the wall face and the near wall node, respectively, and $\overline{q_{\rm e}}$ is the mean thermal speed from Eq.\,\eqref{eqn:qbar}. The Einstein diffusion coefficient $D_{\rm e}$ is given by:
\begin{equation}
D_{\rm e} = \frac{\mu_{\rm e} k_{\rm B} T_{\rm e}}{|C_{\rm e}|}
\label{eqn:Ediffcoeff}
\end{equation}
Solving Eq.\,\eqref{eqn:nonrefbalance} for the face density yields:
\begin{equation}
N_{{\rm e}_0} = \frac{N_{{\rm e}_0}^{\rm nw}}{1 + \dfrac{\overline{q_{\rm e}} \, \Delta x}{8 D_{\rm e}}} \qquad \text{for } E_\eta > 0
\label{eqn:nonrefwall}
\end{equation}
Because this formulation guarantees that $N_{{\rm e}_0}$ at the wall is less than $N_{{\rm e}_0}^{\rm nw}$, the resulting diffusive mass flux at the boundary is strictly outward, perfectly matching the kinetic loss demanded by Eq.\,\eqref{eqn:nonrefbalance}. The boundary conditions for the ions and ballistic groups remain unaffected by this modification.

\section{Numerical Methods}
\label{sec:numerics}

The transport equations are advanced with a finite-volume method on a uniform mesh, the advective fluxes being reconstructed at the cell faces with a total variation diminishing scheme and the limiter of van Leer~\cite{jcp:1979:vanleer}. The limiter is what makes the method usable here: the ballistic groups are strongly directional and their densities vary by many orders of magnitude across a cathode fall, so an unlimited second-order reconstruction produces oscillations and negative densities at the steep fronts, while first-order upwinding smears the very structure the multigroup formulation exists to resolve. Diffusive and conduction fluxes are central, and the chemical source terms are evaluated at the cell centres. Time integration is explicit, save for the turning point of Reaction 6, whose coefficient is large by construction and which is therefore advanced point-implicitly so that its rate does not restrict the step. The breakdown computations, which hinge on whether a transient grows or decays, use a single global time step. The steady cathode sheath of Section~\ref{sec:sheath} is instead marched to convergence with local time stepping, each cell advancing with the step allowed by the stability limits of that cell and its two neighbours; since only the converged solution is sought, this accelerates convergence without affecting the result.

The potential equation is solved semi-implicitly. Rather than relating the potential to the instantaneous space charge alone, the charge that the current will have moved by the end of the step is anticipated, which amounts to adding to the permittivity a term proportional to the local conductivity of the bulk electrons and to the time step. Without it the step is limited by the dielectric relaxation time, which in the dense regions of a discharge is far shorter than any timescale of interest; with it, steps several orders of magnitude larger are stable. Because only the conduction current of the bulk electrons enters it, the added term does not vanish exactly at steady state, and the converged potential departs from Gauss's law by an amount that scales with the ratio of the time step to the local dielectric relaxation time. In the breakdown computations of Section~\ref{sec:breakdown}, the electron densities are low enough for this ratio to remain small, so the added term accelerates the approach to convergence without appreciably altering the converged solution.

The breakdown voltage is obtained by bisection. A supply voltage is proposed, the discharge is marched until it either grows or decays, and the bracket is halved accordingly until the sustaining voltage is located to the required tolerance. Each evaluation must be carried far enough for the ion population to establish itself, and the ion transit time is the slowest scale in the problem by some three orders of magnitude. The ion mobility is therefore multiplied by a thousand for these runs, which compresses that scale without changing the balance that decides breakdown: at threshold the discharge is a Townsend discharge, in which the space charge is too small to distort the field and the criterion depends only on the ionization produced per ion arriving at the cathode and on the secondary emission coefficient, neither of which the mobility enters. The self-sustaining voltage recovered this way is the same as that obtained from a full unaccelerated run at a fraction of the cost. The acceleration is used only to locate the breakdown voltage; the time-dependent profiles of Fig.~\ref{fig:breakdown_ne_eps} are computed with the physical ion mobility.

The same numerical methods are applied to each of the four physical models compared in Section~\ref{sec:tests}. All use the same finite-volume discretization, the same TVD reconstruction and limiter, the same semi-implicit treatment of the potential, and the same bisection procedure for the breakdown voltage, so that the differences reported in the next section are attributable to the physical closure alone and not to the numerics.

\section{Test Cases}
\label{sec:tests}

To assess the proposed multigroup fluid model, we compare its predictions against three established approaches. The first is a kinetic benchmark based on the forward-backward method of Raizer and Shneider \cite{misc:1989:raizer,misc:1991:raizer,pop:2013:macheret}, which resolves the electron distribution in energy and in the two propagation directions rather than representing it by moments, and which serves here as the reference. The second and third are conventional single-fluid models, in which the whole electron population is carried by one drift-diffusion fluid. The second uses the Local Field Approximation (LFA), in which the electron temperature is a function of the local reduced electric field alone. The third uses the Local Energy Approximation (LEA), the extended fluid formulation widely used for glow discharges \cite{pop:2012:rafatov}, in which the transport coefficients depend instead on a transported mean electron energy. The LEA model solves the same electron energy transport equation used for our bulk group, but it omits the ballistic electron groups entirely.

For the Paschen curve test case, each of the four models is solved in conjunction with a potential equation based on Gauss's law. For the cathode-sheath test case, no potential equation is solved, because the potential is prescribed.

All four models share the same heavy-particle physics: the ion transport equations of Section~\ref{sec:heavy} for the breakdown case, and the same frozen ion density for the cathode-sheath case. They also rely on the same argon cross-section dataset from the Hayashi database \cite{lxcat:hayashi}.

We evaluate the models on two problem classes. First, we compute the breakdown voltage in quiescent argon over a range of gas densities to reproduce the Paschen curve. Second, we calculate the current-voltage characteristic of a high-voltage cathode sheath, under conditions typical of discharges in plasma-assisted combustion and plasma aerodynamics.

\subsection{Breakdown Voltage}
\label{sec:breakdown}
\begin{figure}[!t]
     \centering
\includegraphics[width=3.37in]{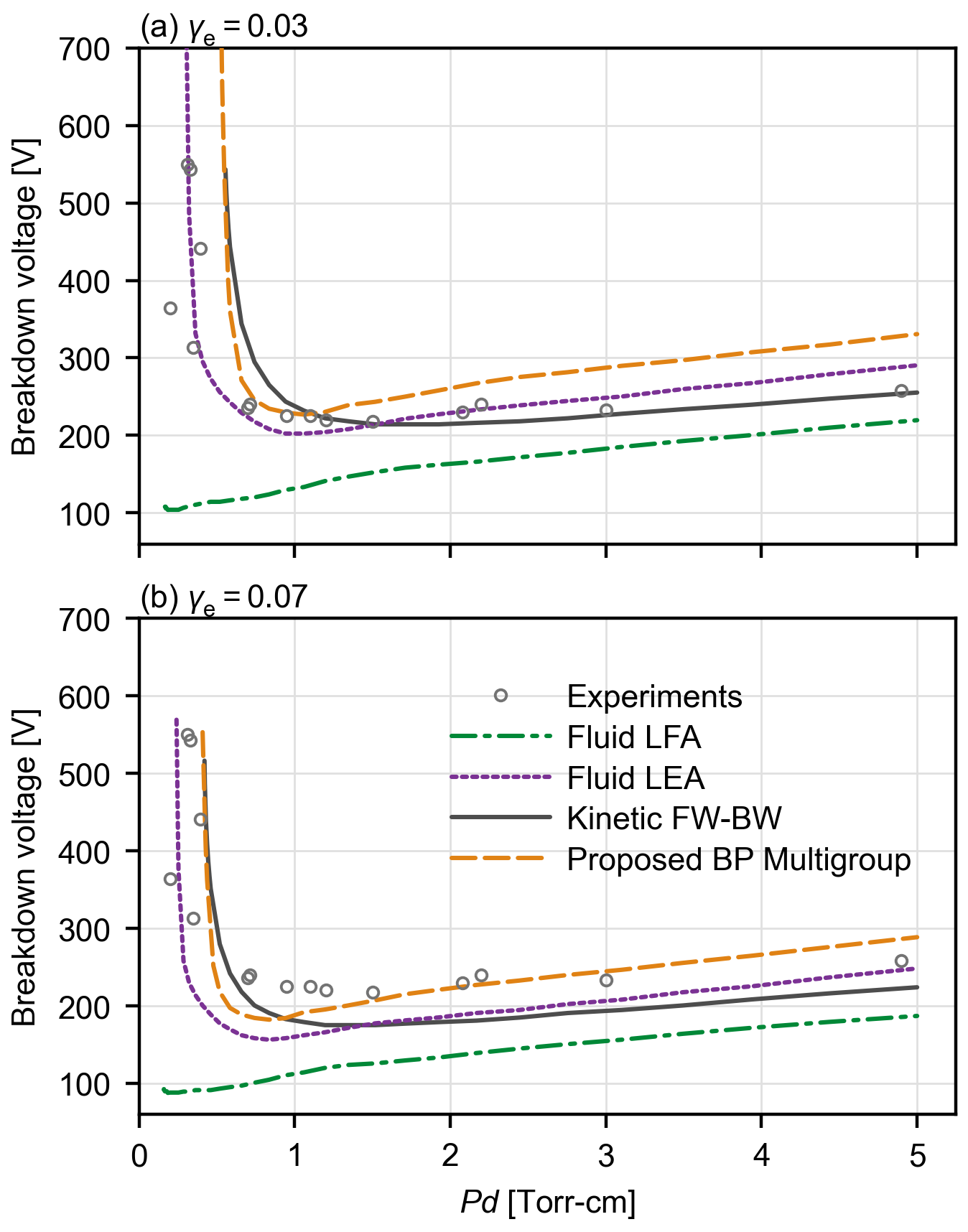}  
     \figurecaption{Breakdown voltage of argon as a function of $Pd$ for a 1~cm gap at 300~K with (a) $\gamma_{\rm e}=0.03$ and (b) $\gamma_{\rm e}=0.07$, predicted by the kinetic forward-backward, LFA fluid, LEA fluid, and proposed multigroup models, together with experimental data \cite{physica:1940:kruithof,prsa:1978:bhasavanich,pre:1997:petrovic,jap:1998:auday,psst:1999:phelps}. Experimental $Pd$ values reported at 0~$^\circ$C are scaled by 300/273 to 300~K.}
     \label{fig:paschen_gammae}
\end{figure}
\begin{figure}[!t]
     \centering
\includegraphics[width=3.37in]{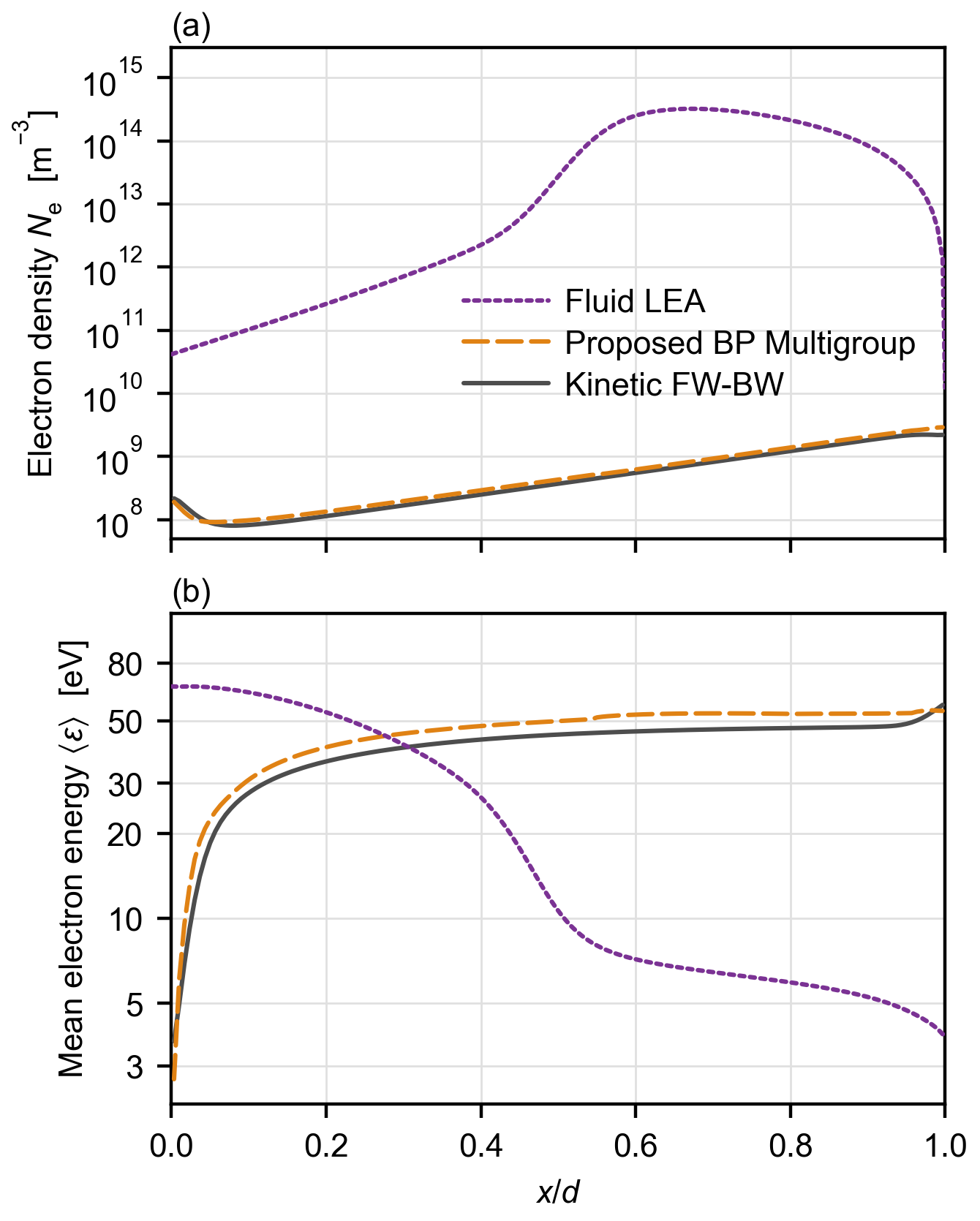}  
     \figurecaption{Electron density $N_{\rm e}$ and mean electron energy $\langle \varepsilon \rangle$ at a time of 10~$\mu$s, $Pd=0.6$~Torr-cm, $\gamma_{\rm e}=0.03$, and a voltage of 500~V for the breakdown case.}
     \label{fig:breakdown_ne_eps}
\end{figure}

The first test case determines the breakdown voltage of argon between two plane-parallel electrodes separated by a gap $d = 1$~cm, with the cathode at $x=0$ and the anode at $x=d$. The gas temperature is fixed at 300~K and the pressure $P$ is varied to sweep the pressure-gap product $Pd$ from 0 to 5~Torr-cm. The secondary emission coefficient is held constant, and two values ($\gamma_{\rm e}=0.03$ and $\gamma_{\rm e}=0.07$) are considered, each yielding its own Paschen curve. At every $Pd$, the breakdown voltage is located by the bisection procedure of Section~\ref{sec:numerics}, between 10 and 600~V and to within 2~V, starting from a uniform seed density of $10^{11}$~m$^{-3}$ and with the ion mobility accelerated as described there. The bulk electrons obey the zero-gradient condition of Eq.~\eqref{eqn:zerograd_e0} at the anode. The gap is discretized with 100 cells, and the multigroup model uses $n_{\rm b}=60$ groups with uniformly spaced levels, with $\phi_{\rm gauge}$ set high enough that the level grid spans the most energetic electrons at every voltage tested. Measured experimental breakdown voltages  are included for reference.

Figure~\ref{fig:paschen_gammae} compares the breakdown voltages predicted by the four models. On the left branch of the Paschen curve, where the breakdown voltage rises steeply as $Pd$ decreases, the multigroup and kinetic curves turn upward at nearly the same $Pd$ and almost overlap. In this regime the reduced field exceeds $E_{\rm crit}$ across the whole gap and a large fraction of the electrons run away. The multigroup model carries these electrons in its ballistic groups and therefore reproduces the $Pd$ below which the discharge can no longer be sustained. The LEA model also produces a left branch, but places it at a lower $Pd$, and the LFA model produces none over the range shown: its breakdown voltage keeps falling as $Pd$ decreases.

Both failures have the same origin. When the reduced field exceeds $E_{\rm crit}$, the electrons run away and cross the gap along nearly straight paths. The number of ionizing collisions such an electron can undergo is then limited by the distance it travels, and raising the voltage does little to increase it, since the ionization cross-section peaks and then falls at high energy. Below some $Pd$, no voltage can provide enough ionization to sustain the discharge, which is what makes the left branch nearly vertical. The drift-diffusion fluids cannot represent this. Their electrons advance slowly across the gap in a random walk, colliding many times on the way, so their ionization keeps rising with the field. Both fluid models therefore overestimate the ionization on the left branch and find breakdown at voltages too low. In the LFA model, where the ionization follows the local reduced field alone, the left branch disappears entirely. The LEA model recovers part of the effect through its transport of electron energy, but still places the left branch at too low a $Pd$.

On the right branch, the reduced field lies below $E_{\rm crit}$, few electrons run away, and the multigroup model is governed by its bulk group. Its curve accordingly follows the trend of the LEA model, lying some 15 to 20\% above it. With the runaway fraction zero throughout the gap, the two models differ only in the treatment of the secondary electrons: in the multigroup model these are emitted into the ballistic group $n_{\rm b}$ and reach the bulk group only once de-beamed, whereas in the LEA model they enter the bulk directly. The offset between the two curves originates from this difference.

The measured experimental breakdown voltages  are shown for reference but are not used as a benchmark here. All the models assume a constant $\gamma_{\rm e}$, whereas the effective emission coefficient of a real cathode depends on its surface condition and, at the high reduced fields of the left branch, on the contributions of energetic ions and neutrals. Phelps and Petrovi\'c \cite{psst:1999:phelps} compiled effective coefficients that rise steeply with the reduced field, but these were not obtained from a kinetic description of the electron avalanche. They were inferred instead by inverting a simplified closed-form expression for the breakdown voltage as a function of $\gamma_{\rm e}$ against the measured voltages. They therefore absorb whatever that expression omits, including the nonlocal electron behavior that is the subject of the present work, and are better regarded as fitting parameters of that expression than as accurate measurements of the emission coefficient. For this reason, the models are compared at the same constant $\gamma_{\rm e}$, and the kinetic solution rather than the experiments serves as the reference.

The shift of the curves with $\gamma_{\rm e}$ provides a consistency check. In classical Townsend theory, both the minimum breakdown voltage and the $Pd$ at which it occurs scale with $\ln(1+1/\gamma_{\rm e})$, which for $\gamma_{\rm e}=0.03$ and 0.07 predicts a ratio of 1.30 between the two curves. The multigroup model gives ratios of 1.25 for the voltage minimum and 1.26 for its $Pd$, and the kinetic model 1.22 and 1.24, within 3\% of the multigroup values. Both models thus depart from the classical ratio by similar amounts, between 3 and 6\%. The agreement between the two models is the more meaningful result, since the classical scaling rests on a local ionization coefficient, which is only approximately valid near the Paschen minimum, where ballistic effects are already significant.

To examine the ballistic regime more closely, Fig.~\ref{fig:breakdown_ne_eps} compares the electron density and mean electron energy on the left branch, at $Pd=0.6$~Torr-cm, $\gamma_{\rm e}=0.03$ and an applied voltage of 500~V, 10~$\mu$s after the discharge is initiated. Because these profiles are snapshots of a transient rather than converged states, they are computed time-accurately, with a single global time step and the ion mobility of Eq.\,\eqref{eqn:muAr} left at its physical value, without the thousandfold acceleration used for the bisection (Section~\ref{sec:numerics}). The mean electron energy $\langle \varepsilon \rangle (x)$ is defined as the local electron energy density divided by the total local electron number density. For the kinetic model, this is the density-weighted average over all forward and backward energy groups. For the multigroup model, it is the density-weighted average over all ballistic groups together with the thermal energy of the bulk group ($\overline{\varepsilon_{\rm e}}$). For the LEA model, it reduces to the bulk thermal energy $\frac{3}{2} k_{\rm B} T_{\rm e}/|C_{\rm e}|$.

The mean energy of the multigroup model follows the kinetic profile closely in shape and lies about 10 to 15\% above it over most of the gap. Both start at a few electron-volts at the cathode, rise over the first two millimetres as the emitted and newly liberated electrons are accelerated, and level off across the rest of the gap, near 47~eV in the kinetic model and 53~eV in the multigroup model, where the energy gained from the field is offset on average by inelastic losses and by the continual addition of cold electrons from ionization. The LEA model behaves in the opposite way: its mean energy is highest at the cathode, about 65~eV, and decays across the gap to a few electron-volts. Having no inertia, the LEA electrons take on the energy dictated by the local energy balance the moment they are emitted, so the acceleration region near the cathode is absent, and the energetic, forward-directed electrons that carry the ionization across the gap cannot be represented.

The electron densities in Fig.~\ref{fig:breakdown_ne_eps}a agree as closely. The multigroup density lies about 15\% above the kinetic one over most of the gap, and the two grow from cathode to anode at nearly the same exponential rate, their slopes on the logarithmic scale being almost identical. The three models are not, however, equally far above breakdown. At $Pd=0.6$~Torr-cm, the breakdown voltages are 420~V (kinetic), 330~V (multigroup), and 240~V (LEA). Although the multigroup and kinetic left branches nearly coincide in $Pd$, the branch is steep enough here that a small shift in $Pd$ translates into a large difference in voltage. An applied voltage of 500~V thus exceeds the breakdown voltage by about 20\% in the kinetic model, 50\% in the multigroup model, and 110\% in the LEA model. Because these profiles are growing avalanches rather than steady states, the least overvolted kinetic avalanche has the lowest density at this instant, followed by the multigroup and then the LEA avalanche. The LEA density, however, departs from the other two in structure as well as in magnitude: it exceeds them by two to nearly six orders of magnitude, grows much faster, peaks about two-thirds of the way across the gap, and falls toward the anode, a fundamentally different profile that reflects its inability to capture ballistic transport.

\begin{figure}[!t]
     \centering
\includegraphics[width=3.37in]{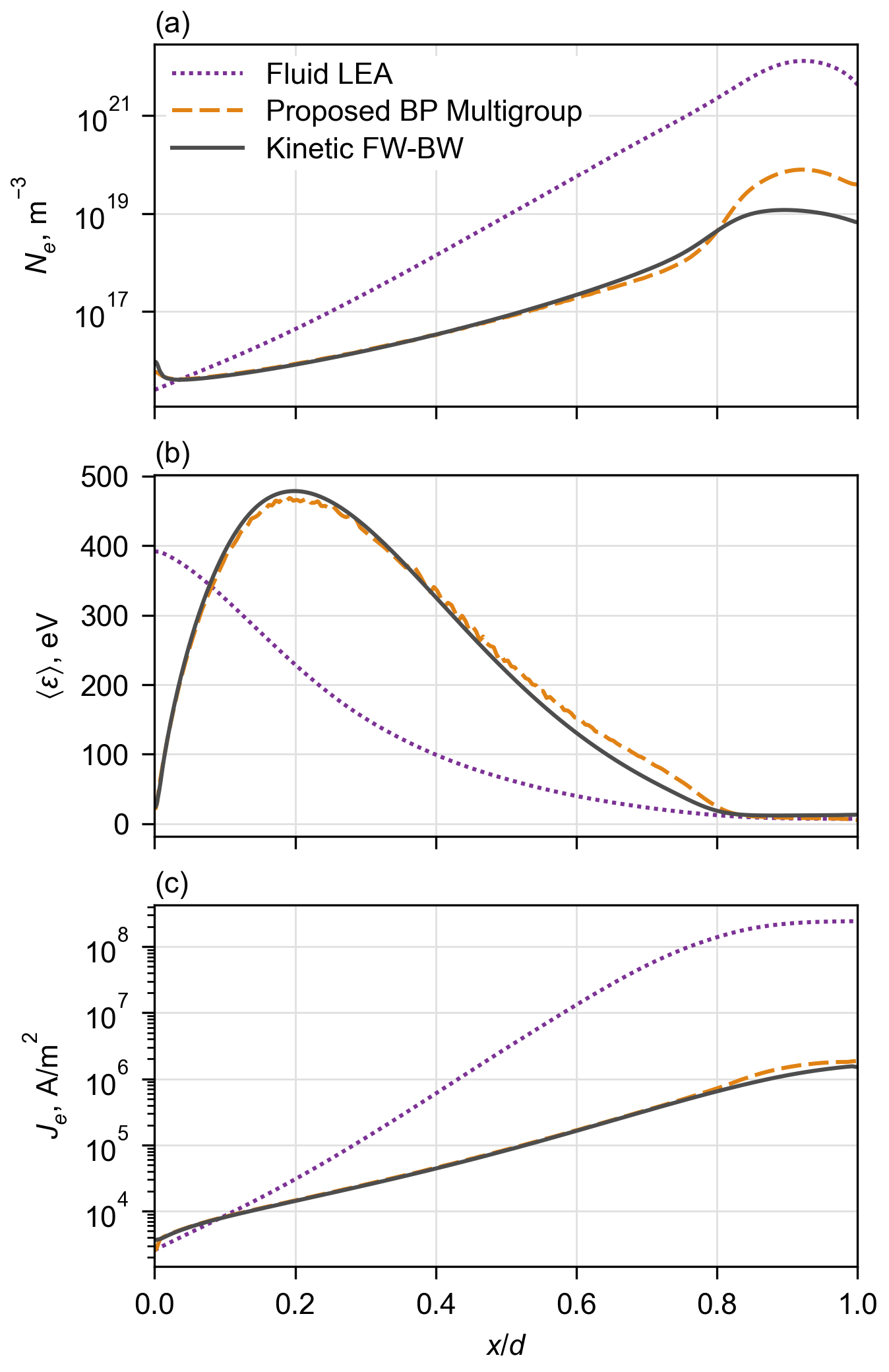}  
     \figurecaption{(a) Electron number density $N_{\rm e}$, (b) mean electron energy $\langle \varepsilon \rangle$, and (c) electron current density $J_{\rm e}$ across the prescribed-potential cathode sheath at $V_{\rm c}=3000$~V, for argon at 0.25~atm and 1500~K with a frozen ion density $N_{\rm Ar^+}=10^{19}$~m$^{-3}$ and $\gamma_{\rm e}=0.3$.}
     \label{fig:sheath-Ne-eps-Je}
\end{figure}
\begin{figure}[!t]
     \centering
\includegraphics[width=3.37in]{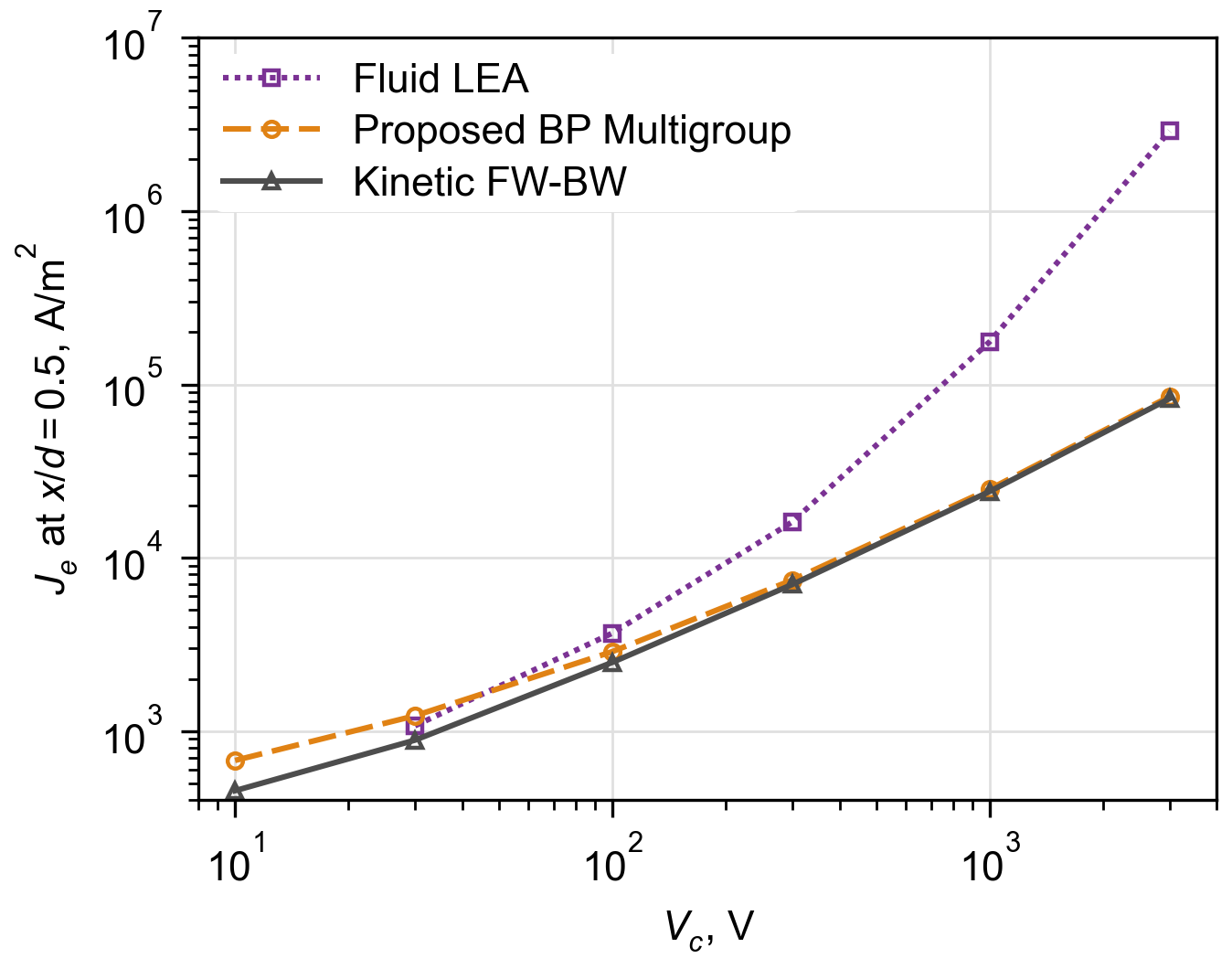}  
     \figurecaption{Electron current density $J_{\rm e}$ at mid-sheath ($x/d=0.5$) as a function of the cathode-fall voltage $V_{\rm c}$ for the prescribed-potential cathode sheath of Fig.~\ref{fig:sheath-Ne-eps-Je}.}
     \label{fig:sheath-current-voltage}
\end{figure}

\subsection{Cathode Sheath}
\label{sec:sheath}

The second test case is a high-voltage cathode sheath in argon at 0.25~atm and 1500~K ($N_{\rm Ar}=1.22\times10^{24}$~m$^{-3}$), conditions representative of plasma-assisted combustion and supersonic combustion applications. To compare the electron transport of the models in isolation, we adopt the standard prescribed-field sheath configuration: the ion density is held fixed in space and time at $N_{\rm Ar^+}=10^{19}$~m$^{-3}$, and the potential is imposed rather than obtained from Gauss's law. It rises from the cathode at $x=0$ to the sheath edge at $x=d$ as
\begin{equation}
\phi(x) = V_{\rm c}\left[1-\left(1-\frac{x}{d}\right)^{3}\right]
\label{eqn:sheathphi}
\end{equation}
where $V_{\rm c}$ is the cathode-fall voltage. The corresponding field, $E=-3V_{\rm c}(1-x/d)^2/d$, is strongest at the cathode and vanishes at the sheath edge. Gauss's law applied to Eq.~\eqref{eqn:sheathphi} gives a net charge density of $6\epsilon_0 V_{\rm c}(1-x/d)/d^2$, which falls linearly to zero at the sheath edge, where the plasma becomes quasi-neutral. The sheath thickness is fixed by requiring this charge density to equal $|C_{\rm e}|N_{\rm Ar^+}$ at the cathode:
\begin{equation}
d = \sqrt{\frac{6\,\epsilon_0 V_{\rm c}}{|C_{\rm e}|\,N_{\rm Ar^+}}}
\label{eqn:sheathd}
\end{equation}
so that $d$ grows as $V_{\rm c}^{1/2}$, from 18~$\mu$m at $V_{\rm c}=10$~V to 315~$\mu$m at $V_{\rm c}=3000$~V. The imposed potential is thus consistent with Gauss's law if the electron density rises linearly from zero at the cathode to $N_{\rm Ar^+}$ at the sheath edge; it is not adjusted to the electron density that each model actually predicts. Freezing the ions and the potential removes the coupling between the electrons and the space charge. In a self-consistent calculation, any error in the electron multiplication would alter the ion density, which would in turn reshape the sheath field and change the ion flux to the cathode and hence the secondary emission. These effects compound one another, so the differences between the models would be amplified and difficult to attribute to any single cause. With the ions and the field fixed, the ion flux to the cathode, and therefore the emitted electron current, is identical in all the models, and every difference in the electron current downstream results from the electron transport alone. Secondary electrons are emitted with $\gamma_{\rm e}=0.3$ and $\varepsilon_{\rm emit}=1$~eV, and the sheath edge is treated as a perfectly absorbing anode (Section~\ref{sec:nonreflecting}). The sheath is discretized with 100 cells, and the multigroup model uses $n_{\rm b}=200$ groups with uniformly spaced levels. At $V_{\rm c}=3000$~V, the reduced field at the cathode is about 23\,000~Td ($2.3\times10^{-17}$~V$\cdot$m$^2$), roughly 40 times the critical field of Eq.~\eqref{eqn:Ecrit} ($E_{\rm crit}/N_{\rm Ar}\approx 590$~Td), and the field exceeds $E_{\rm crit}$ over about the first 85\% of the sheath.

Figure~\ref{fig:sheath-Ne-eps-Je} shows the electron density, mean electron energy and electron current density across the sheath at $V_{\rm c}=3000$~V. All three models start from the same electron current at the cathode, set by the emission, so the differences that develop across the sheath reflect electron multiplication alone. The kinetic and multigroup models agree closely throughout. In both, the mean energy starts near the 1~eV emission energy, climbs to a maximum of about 480~eV near $x/d\approx 0.2$ as the emitted electrons are accelerated by the strong field, and then declines as the field weakens and as the cold electrons liberated by ionization come to outnumber the energetic ones. Beyond $x/d\approx 0.8$, where the field has fallen below $E_{\rm crit}$, the population is dominated by thermal electrons. The LEA model, in contrast, assigns its largest energy, about 400~eV, to the cathode itself and decreases monotonically from there. Lacking inertia, its electrons take on the local-equilibrium energy immediately upon emission, and the acceleration region adjacent to the cathode, where the emitted electrons have not yet gained enough energy to ionize efficiently, is absent.

The consequences for the current are large. Across the sheath, the LEA current grows by more than four orders of magnitude, compared with about two and a half for the kinetic and multigroup models, so that the LEA model overpredicts the current reaching the sheath edge by roughly two orders of magnitude. The cause is the same as on the left branch of the Paschen curve. A drift-diffusion model assumes that the energy an electron gains from the field is balanced locally by its collisional losses, and describes its motion as a slow drift superposed on a random walk. Above $E_{\rm crit}$ no such balance exists: the electrons run away and cross the sheath on nearly straight paths. The LEA electrons instead advance at the much slower drift velocity and so undergo far more ionizing collisions per unit distance than the runaway electrons they are meant to represent. The density profiles reflect the same behavior. Toward the sheath edge the field vanishes, the electrons slow down, and the density rises sharply in all three models. The LEA density reaches about $10^{21}$~m$^{-3}$ there, two orders of magnitude above the ion density; were the potential computed self-consistently, a space charge of this magnitude would be incompatible with the imposed sheath and would collapse it. The multigroup density peaks close to the ion density, and the kinetic density a few times below it. This residual difference is confined to the weak-field region near the sheath edge, where the electrons are no longer ballistic and are carried by the bulk group of the multigroup model. It depends strongly on the boundary condition imposed there. With the zero-gradient anode condition of Eq.\,\eqref{eqn:zerograd_e0}, the multigroup density near the sheath edge exceeds the kinetic one by more than an order of magnitude; the absorbing wall of Section~\ref{sec:nonreflecting} reduces this excess several-fold, to the factor of about 4 seen in Fig.~\ref{fig:sheath-Ne-eps-Je}(a). In both cases the density over most of the sheath is unaffected. The discrepancy therefore arises largely from the boundary treatment rather than from the model at the interior nodes, although the local closure of the bulk group, which inherits there in much reduced form the limitations of the LEA model, may account for part of what remains.
\begin{figure*}[t]
     \centering
\includegraphics[width=6.69in]{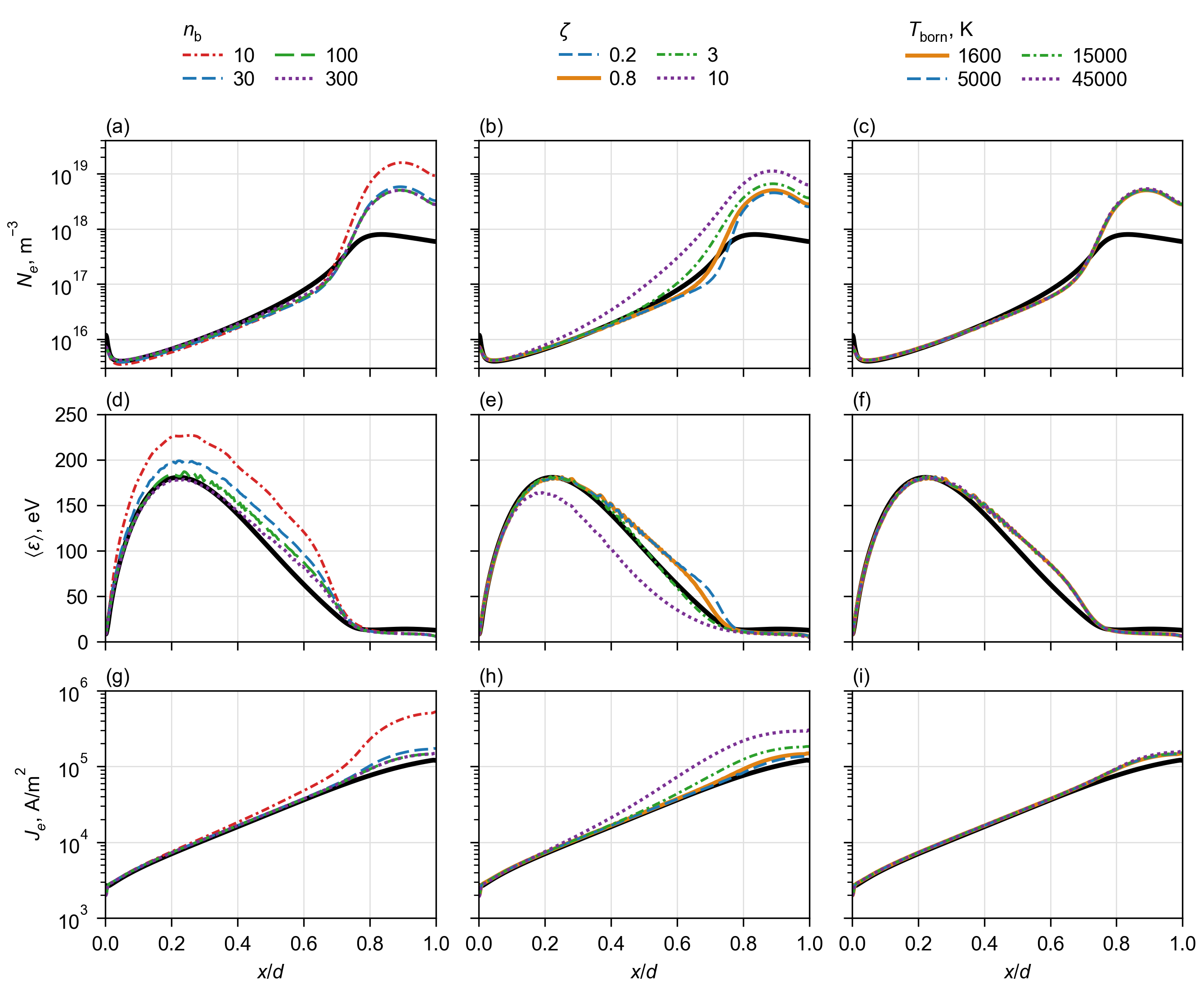}  
     \figurecaption{Sensitivity of the prescribed-potential cathode sheath of Fig.~\ref{fig:sheath-Ne-eps-Je} at $V_{\rm c}=1000$~V to (a,d,g) the number of groups $n_{\rm b}$, (b,e,h) the straggling parameter $\zeta$, and (c,f,i) the birth temperature $T_{\rm born}$: (a--c) electron number density $N_{\rm e}$, (d--f) mean electron energy $\langle \varepsilon \rangle$, and (g--i) electron current density $J_{\rm e}$. Parameters not varied are held at their baseline values; the kinetic forward-backward solution is shown for reference using a thick black line.}
     \label{fig:sensitivity-zeta-nb-Tborn}
\end{figure*}
\begin{figure}[t]
     \centering
\includegraphics[width=3.37in]{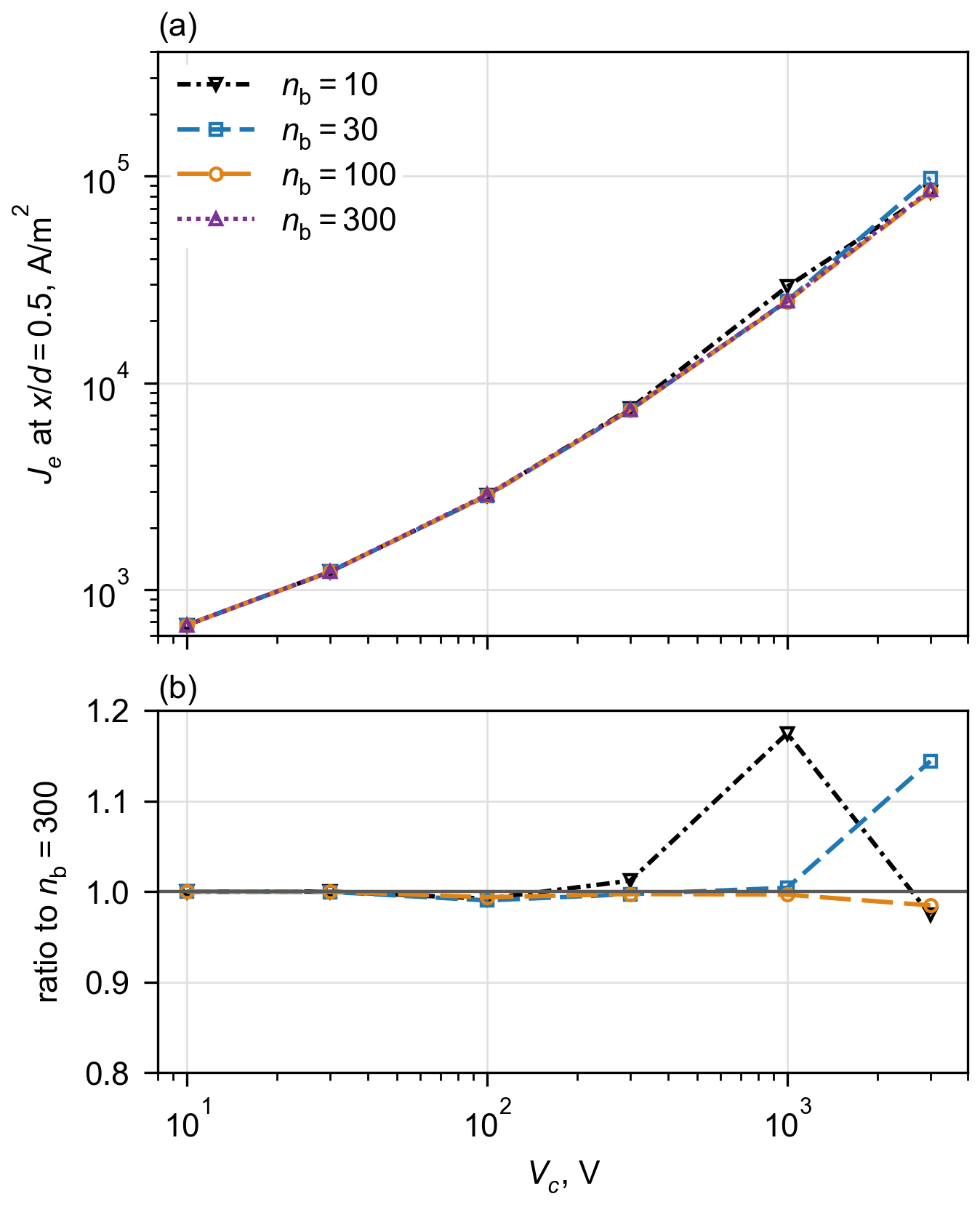}  
     \figurecaption{(a) Electron current density $J_{\rm e}$ at mid-sheath ($x/d=0.5$) as a function of the cathode-fall voltage for several numbers of groups, and (b) the same normalized by the $n_{\rm b}=300$ result, for the prescribed-potential cathode sheath of Fig.~\ref{fig:sheath-Ne-eps-Je}.}
     \label{fig:sensitivity-current-voltage}
\end{figure}

Figure~\ref{fig:sheath-current-voltage} shows how the electron current density at mid-sheath, $x/d=0.5$, varies with the cathode-fall voltage from 10 to 3000~V. Over this range, the sheath thickness increases from 18 to 315~$\mu$m and the reduced field at the cathode from about 1350 to 23\,000~Td, that is, from about 2 to 40 times $E_{\rm crit}$, so that the runaway region grows from about a third of the sheath to most of it. These results are converged in the number of groups $n_{\rm b}$, as shown in Section~\ref{sec:sensitivity}. At the lowest voltages, the models give comparable currents. With a cathode fall of a few tens of volts, the electrons gain too little energy to ionize much, and the current at mid-sheath is set by the emission, which is identical in all models, less the electrons scattered back to the cathode. The multigroup current nonetheless exceeds the kinetic one by about 45\% at $V_{\rm c}=10$~V and 40\% at 30~V. At these voltages, however, the kinetic model is not necessarily the better reference. The current is then set almost entirely by how many emitted electrons are scattered back to the cathode, and the forward-backward model, which restricts the electrons to two directions of motion, represents this angular redistribution only approximately. Low-voltage collisional sheaths of this kind are commonly computed with the LEA model \cite{pop:2012:rafatov}, and at 30~V the LEA current lies roughly midway between the other two, about 20\% above the kinetic value and 15\% below the multigroup one. The three models thus differ by comparable amounts, and none can be singled out as the most accurate. On the multigroup side, the likely source of the difference is the local runaway partition, which is least accurate in this regime (Section~\ref{sec:runaway}): the field exceeds $E_{\rm crit}$ near the cathode, but the total fall is far smaller than the climb of about 75~V to the friction peak, so the partition returns most elastically scattered electrons to the beam, some of which would in fact be scattered back to the cathode. The differences diminish steadily as $V_{\rm c}$ increases, and from about 300~V upward, where the electrons are ballistic over most of the sheath and the kinetic model is most reliable, the multigroup current agrees with the kinetic one to within a few percent.

The LEA model departs from the kinetic model in the opposite direction, and increasingly so with voltage. Its current exceeds the kinetic one by a factor of about 2 at 300~V, 8 at 1000~V and 30 at 3000~V. As $V_{\rm c}$ increases, the runaway region extends over a larger part of the sheath, so the excess ionization per unit length of the LEA model acts over more of it; because multiplication depends exponentially on the integrated ionization, the error compounds. The multigroup model, which carries the runaway electrons in its ballistic groups, tracks the kinetic current-voltage characteristic across the full range, and its discrepancy is smallest precisely at the high voltages where that of the LEA model is largest.

\subsection{Sensitivity to Adjustable Parameters}
\label{sec:sensitivity}

Besides the cross-section data, the multigroup model depends on a few parameters that the physics does not fix: the number of groups $n_{\rm b}$, which sets the resolution of the birth-potential axis; the straggling parameter $\zeta$ of the runaway partition; and the birth temperature $T_{\rm born}$ at which volumetrically liberated electrons are created. Their influence is assessed on the prescribed-potential cathode sheath of Section~\ref{sec:sheath}. Figure~\ref{fig:sensitivity-zeta-nb-Tborn} shows the profiles at $V_{\rm c}=1000$~V. At this voltage the sheath is 182~$\mu$m thick, the reduced field at the cathode is about 13\,000~Td, some 22 times $E_{\rm crit}$, and runaway is possible over the first 80\% of the sheath. Each parameter is varied in turn with the others held at their baseline values, $n_{\rm b}=150$, $\zeta=0.8$ and $T_{\rm born}=1600$~K. Figure~\ref{fig:sensitivity-current-voltage} complements this with the mid-sheath current over the full voltage range for several values of $n_{\rm b}$.

The number of groups is a numerical rather than a physical parameter, and the solution should converge as it increases. Since $\phi_{\rm gauge}$ is set just above the potential of the sheath edge, the $n_{\rm b}-1$ volumetric levels span approximately the cathode fall, and each is about $V_{\rm c}/(n_{\rm b}-1)$ wide. At 1000~V this width ranges from about 110~V for $n_{\rm b}=10$ to about 3~V for $n_{\rm b}=300$. The profiles for $n_{\rm b}=30$, 100 and 300 in Fig.~\ref{fig:sensitivity-zeta-nb-Tborn}(a,d,g) are nearly indistinguishable, and the peak mean energy converges onto the kinetic value from above as the levels are refined. Only the coarsest grid departs markedly. With levels about 110~V wide, a group's energy is resolved no more finely than the range over which the ionization cross-section rises to its peak. As a result, $n_{\rm b}=10$ overestimates the peak mean energy by about 25\% and the current reaching the sheath edge by a factor of about 3. This large error is confined to the near-anode region adjacent to the sheath edge, which is considerably affected by the anode boundary condition of Section~\ref{sec:nonreflecting}; elsewhere in the sheath, the error in the current remains below 10\%.

The mid-sheath current of Fig.~\ref{fig:sensitivity-current-voltage} shows the same behavior over the whole voltage range. For $n_{\rm b}\geq100$ it lies within about 5\% of the $n_{\rm b}=300$ result at every voltage. For $n_{\rm b}=30$ it stays within 5\% up to 1000~V but is 12\% high at 3000~V. For $n_{\rm b}=10$ the error reaches 18\% at 1000~V. Thus, as few as 10 to 30 groups, depending on the voltage, predict the current to within about 12\% over the whole range of cathode-fall voltages considered herein. This accuracy is sufficient for the intended application: the numerical errors incurred when simulating complex problems with computational fluid dynamics solvers often exceed 10\%, so resolving the current more finely brings little practical benefit. The error is governed by the level width rather than by $n_{\rm b}$ itself. Widths of about 30~V, reached for $n_{\rm b}=10$ at 300~V, $n_{\rm b}=30$ at 1000~V and $n_{\rm b}=100$ at 3000~V, keep it to 2 or 3\%, whereas widths near 100~V raise it to 12 to 18\%. The required number of groups thus scales with the voltage spanned by the level grid and not with the number of cells. The $n_{\rm b}=200$ used in Section~\ref{sec:sheath}, with levels about 15~V wide at 3000~V, lies well within the converged range. The coarse-grid errors are larger at the sheath edge than at mid-sheath (compare Fig.~\ref{fig:sensitivity-zeta-nb-Tborn}(g) with Fig.~\ref{fig:sensitivity-current-voltage}), but from $n_{\rm b}=30$ upward the current is converged across the whole sheath.

The straggling parameter was set to $\zeta=0.8$ because this value gave the closest agreement with the kinetic electron current over all the cathode-fall voltages considered in this sheath case. The question is therefore how sharply that agreement depends on it, and Fig.~\ref{fig:sensitivity-zeta-nb-Tborn}(b,e,h) shows that it depends on it only weakly. The profiles for $\zeta=0.2$, 0.8 and 3 nearly coincide with one another and, in mean energy, with the kinetic solution. Only $\zeta=10$ departs noticeably: its mean energy peaks about 10\% lower and collapses to thermal values by $x/d\approx0.6$ rather than 0.8, and its sheath-edge current is about 60\% above the baseline. Equation~\eqref{eqn:frunstep} explains this. At $|E|=5E_{\rm crit}$ the runaway fraction is 0.99 for $\zeta=0.8$, 0.74 for $\zeta=3$ and 0.33 for $\zeta=10$, and for $\zeta\leq0.8$ it exceeds 0.9 wherever $|E|>3E_{\rm crit}$. Since the field lies far above this over most of the sheath, $\zeta$ matters only in the outer part, where the field falls toward $E_{\rm crit}$, unless it is large enough to suppress runaway even in strong fields. A larger $\zeta$ keeps more of the liberated electrons in the bulk group, where they multiply as in the LEA model, which is why the current rises monotonically with $\zeta$. The calibration therefore constrains $\zeta$ only loosely. Values from 0.2 to 3 change the sheath-edge current by no more than about 20\%, so the agreement of Fig.~\ref{fig:sheath-current-voltage} does not rest on fine-tuning. The breakdown voltages of Section~\ref{sec:breakdown} are less sensitive still to $\zeta$, and so provide a test of the model that is largely independent of this calibration.

The birth temperature has the least influence of the three. Raising it from 1600 to 45\,000~K increases the creation energy $\frac{3}{2}k_{\rm B}T_{\rm born}$ from 0.2 to 5.8~eV. This leaves the density and mean-energy profiles of Fig.~\ref{fig:sensitivity-zeta-nb-Tborn}(c,f) nearly unchanged and raises the current at the sheath edge by about 20 to 30\%. $T_{\rm born}$ does not enter the ballistic groups at all, since a volumetrically liberated electron joins them with zero forward energy whatever its temperature (Eq.~\eqref{eqn:phibirth}). It acts only through the bulk group: through the energy $Q^{\rm born}_{{\rm e}_0}$ credited at creation (Eq.~\eqref{eqn:Qborn}), and through the turning-point rate coefficient of Eq.~\eqref{eqn:turnrate}, which transfers decelerating groups to the bulk earlier when $\varepsilon_{\rm born}$ is larger. Both routes warm the bulk in the weak-field region near the sheath edge and so raise the bulk ionization there, which accounts for the modest rise in current.

One feature is unaffected by all three parameters: near the sheath edge, the electron density of the multigroup model remains well above the kinetic one. Because neither the resolution of the ballistic groups nor the runaway partition nor the creation energy moves it, the excess does not originate in the treatment of the ballistic electrons, consistent with its attribution in Section~\ref{sec:sheath} to the anode boundary condition.

\section{Conclusions}
\label{sec:conclusions}

This paper introduces, to our knowledge for the first time, a multi-fluid electron model in which electron groups are defined by the electrostatic potential at their point of creation. Because the birth potential is a constant of ballistic motion, the electric field does not shift electrons between groups; only collisions do. A group collects every electron liberated within its birth-potential window, regardless of where it is liberated. This approach resolves the primary difficulties of earlier multibeam and multigroup fluid models. Schemes that launch a separate beam from each birth location tie groups to mesh points, so the number of groups scales with the grid. Schemes that bin electrons by kinetic energy must transfer electrons between groups continuously as the field accelerates them. The present model instead uses a fixed number of groups, set solely by the desired energy resolution and independent of the mesh.

This resolution requirement is modest. Depending on the cathode fall, 10 to 30 groups reproduce the current-voltage characteristic of a high-voltage argon cathode sheath to within roughly 10\% of the converged result, comparable to the numerical error of a typical computational fluid dynamics (CFD) simulation. With a converged number of groups, the model tracks the kinetic solution to within a few percent at reduced fields exceeding $10^4$~Td, where the local energy approximation (LEA) overpredicts the current up to thirtyfold. On the left branch of the Paschen curve, where no sheath exists and ballistic electrons originate in the volume, the model recovers the pressure-gap product below which breakdown cannot be sustained, a limit that the local field approximation (LFA) misses entirely and the LEA misplaces.

Consequently, this model is highly compatible with multi-species CFD solvers. Each electron group acts as an additional species governed by a single continuity equation. Its velocity is determined algebraically from the local potential, and its source terms are simply local entries in the reaction mechanism. The groups are advanced using the same spatial discretization and time integration as other species, with losses entering the Jacobian as stabilizing diagonal terms. Adding 10 to 30 such species achieves near-kinetic accuracy in high-voltage cathode sheaths without the computational bottlenecks that have traditionally excluded kinetic methods from fluid codes. Specifically, it avoids the additional velocity-space dimensions required by Boltzmann solvers and the statistical noise of particle methods, which slows convergence to a steady state and complicates fluid coupling.

Extension to two and three dimensions is straightforward. While the birth potential dictates a group's speed, its direction can be approximated by that of the bulk electron velocity. Because the ballistic groups are significant only where the electric field exceeds $E_{\rm crit}$, the bulk velocity in these regions is drift-dominated and closely aligned with the electric force accelerating the ballistic electrons. Thus, no angular coordinate is required. This approximation is least accurate where fast electrons penetrate via inertia into the weak fields beyond the sheath edge, such as in the negative glow. In such regions, this approach should eventually be validated against multidimensional kinetic solutions.

A fundamental requirement of this formulation is that the local potential along an electron's path must change by much less than one level width during its transit. It is the local potential that governs this condition, rather than the electrode potential, as ion space charge can dynamically reshape the sheath. This condition is readily satisfied in DC discharges, where an electron crosses the sheath in roughly a nanosecond while the sheath itself evolves on the much slower ion transit timescale. As a result, the model is also well-suited for pulsed-DC and low-frequency excitation. Radio-frequency discharges at 13.56~MHz lie near the limit of this approximation and represent a promising, albeit marginal, application that may require corrections for the time-varying potential.

The remaining discrepancies with kinetic solutions are of two kinds. At cathode falls of a few tens of volts, the electrons are no longer ballistic, and in this collision-dominated regime a local fluid description such as the bulk group is not expected to incur large errors. Near the anode, the electron density exceeds the kinetic one by an amount that depends strongly on the anode boundary condition but leaves the density over the rest of the domain unaffected, which points to the boundary treatment rather than to the model itself. Thus, this multigroup model retains the efficiency and accuracy of conventional fluid models where electrons are thermal, while extending that accuracy at a modest computational cost into regimes where they run away.

\appendix
\section{Mean Group Potential Offset $\xi_j(\phi)$}
\label{sec:appA}

This appendix provides the evaluation of the dimensionless offset $\xi_j(\phi)$ that defines the mean group potential $\overline{\phi}_j(\phi)$ in Eq.\,\eqref{eqn:phibar}, starting from the centroid of Eq.\,\eqref{eqn:centroid_def}. Defining the normalized birth coordinate $u$ and cutoff coordinate $y$,
\begin{equation}
u = \frac{\phi_{\rm birth} - \phi^{\rm mid}_j}{\Delta \phi_j}, \qquad y = \frac{\phi - \phi^{\rm mid}_j}{\Delta \phi_j}
\label{eqn:app_uy_def}
\end{equation}
the group bounds $\bigl[\phi^{\rm lo}_j, \phi^{\rm hi}_j\bigr]$ map to $u \in \bigl[-\tfrac{1}{2}, \tfrac{1}{2}\bigr]$. Because no volumetric mass is delivered below the lower bound ($g_j = 0$ for $u < -\tfrac{1}{2}$), the dimensionless displacement $\xi_j(\phi) = \bigl(\overline{\phi}_j(\phi) - \phi^{\rm mid}_j\bigr)/\Delta\phi_j$ reduces to:
\begin{equation}
\xi_j(\phi) = \frac{\displaystyle \int_{-1/2}^{y} u \, g_j(u) \, {\rm d}u}{\displaystyle \int_{-1/2}^{y} g_j(u) \, {\rm d}u}
\label{eqn:app_Goffset}
\end{equation}

Equation\,\eqref{eqn:kernel} partitions runaway electrons between the bracketing group $j_{\rm A}$ (share $1-s$) and the adjacent more energetic group $j_{\rm A}+1$ (share $s$). Group $j$ therefore receives mass from births in its own window $\bigl[-\tfrac{1}{2}, \tfrac{1}{2}\bigr]$ ($j_{\rm A} = j$, where $s = \tfrac{1}{2}-u$), and from births in the adjacent lower-energy window $j-1$ ($j_{\rm A} = j-1$). Because electrostatic potential decreases with increasing energy (Eq.\,\eqref{eqn:ev2volt}), window $j-1$ spans $\tfrac{1}{2} \le u \le \tfrac{1}{2} + r$ in units of $\Delta\phi_j$, where $r = \Delta\phi_{j-1}/\Delta\phi_j$. In that window, $s = 1 - \bigl(u - \tfrac{1}{2}\bigr)/r$.

For standard volumetric groups ($j < n_{\rm b}-1$), the share from its own window is $1-s = \tfrac{1}{2}+u$. For the most energetic volumetric group ($j = n_{\rm b}-1$), the emission group $n_{\rm b}$ receives no volumetric mass, so the boundary share $s$ is retained entirely by group $n_{\rm b}-1$, yielding $(1-s)+s = 1$ across its primary window. Thus, in terms of $u$:
\begin{equation}
g_j(u) = \begin{cases}
 \tfrac{1}{2} + u, & -\tfrac{1}{2} \le u \le \tfrac{1}{2}, \quad j < n_{\rm b}-1 \\[1.2ex]
 1, & -\tfrac{1}{2} \le u \le \tfrac{1}{2}, \quad j = n_{\rm b}-1 \\[1.2ex]
 1 - \dfrac{u - \tfrac{1}{2}}{r}, & \phantom{-}\tfrac{1}{2} \le u \le \tfrac{1}{2} + r \\[2.2ex]
 0, & \text{otherwise}
 \end{cases}
\label{eqn:app_deposit}
\end{equation}

Substituting Eq.\,\eqref{eqn:app_deposit} into Eq.\,\eqref{eqn:app_Goffset} and integrating piecewise over the operational intervals yields the closed-form offsets. For standard volumetric groups ($j < n_{\rm b}-1$), evaluating the integrals and introducing the shifted coordinate $z = y - \tfrac{1}{2}$ across the adjacent window gives:
\begin{equation}
\xi_j(\phi) = \begin{cases}
 -\tfrac{1}{2}, & y \le -\tfrac{1}{2} \\[1.5ex]
 \dfrac{4y-1}{6}, & -\tfrac{1}{2} \le y \le \tfrac{1}{2} \\[2.5ex]
 \dfrac{r + 6rz + 6rz^2 - 3z^2 - 4z^3}{6 \left( r + 2rz - z^2 \right)}, & 0 \le z \le r \\[2.5ex]
 \dfrac{1 + 2r}{6}, & z \ge r
 \end{cases}
\label{eqn:Gclosed}
\end{equation}
where $r = \Delta\phi_{j-1}/\Delta\phi_j$, $y = \bigl(\phi - \phi^{\rm mid}_j\bigr)/\Delta\phi_j$, and $0 \le z \le r$ corresponds to $\tfrac{1}{2} \le y \le \tfrac{1}{2}+r$. The least energetic group ($j = 1$) requires no separate treatment: it has no lower-energy window $j-1$, and when the level list starts at $\varepsilon^{\rm lo}_1 = 0$ its upper bound coincides with $\phi_{\rm gauge}$, which the local potential does not exceed, so $r_1$ is never evaluated.

For the most energetic volumetric group ($j = n_{\rm b}-1$), the uniform deposit $g = 1$ across its own window yields:
\begin{equation}
\xi_{n_{\rm b}-1}(\phi) = \begin{cases}
 -\tfrac{1}{2}, & y \le -\tfrac{1}{2} \\[1.5ex]
 \dfrac{2y-1}{4}, & -\tfrac{1}{2} \le y \le \tfrac{1}{2} \\[2.5ex]
 \dfrac{6rz + 6rz^2 - 3z^2 - 4z^3}{6 \left( 2r + 2rz - z^2 \right)}, & 0 \le z \le r \\[2.5ex]
 \dfrac{r(3+2r)}{6(2+r)}, & z \ge r
 \end{cases}
\label{eqn:Gdeep}
\end{equation}
\section*{Data Availability}

The data that support the findings of this study are available from the corresponding author upon reasonable request.

\section*{AI Assistance}

Artificial intelligence tools were utilized to streamline the preparation of this work. Specifically, Claude Opus 5 (Anthropic; Claude Code, June--September 2026) and Gemini 3.8 Flash (Google; September 2026) were employed to reduce manual algebraic and implementation effort across four primary tasks: (i) deriving closed-form expressions for the integrals in Eqs.\,(\ref{eqn:Gclosed})--(\ref{eqn:Gdeep}) from the definition outlined in Eq.\,(\ref{eqn:app_Goffset}); (ii) determining optimal spline fits---achieving a targeted accuracy with the minimum number of nodes---for the reduced electron mobility, reduced electric field, and ionization rates presented in Tables~\ref{tab:ei_splines_Ar}, \ref{tab:mueN}, and \ref{tab:Estar}; (iii) translating the model equations of Secs.~\ref{sec:ballistic}--\ref{sec:heavy} into the C implementation used for all numerical results; and (iv) writing the data processing scripts used to render Figs.~\ref{fig:paschen_gammae}--\ref{fig:sensitivity-current-voltage} from the computational output.

All analytical results generated with AI assistance were independently verified by the authors through symbolic differentiation and by comparison with numerical quadrature, confirming accuracy to a relative tolerance of $10^{-10}$ over the full reported parameter range. Furthermore, all figures present unmodified computed output. The physical model, its theoretical derivation, the interpretation of results, and all conclusions are the authors' own, and the authors take full responsibility for the accuracy and integrity of this work.

\section*{Acknowledgements}

The authors are thankful to the National Science Foundation Transport Phenomena program NSF 26-518 for supporting this research under grant \#2607765.

\footnotesize
\bibliography{all}

@article{pf:2024:parent,
  title     = {Progress in electron energy modeling for plasma flows and discharges},
  author    = {Parent, Bernard and Rodriguez Fuentes, Felipe Martin},
  journal   = {Physics of Fluids},
  volume    = {36},
  number    = {8},
  pages     = {086113},
  year      = {2024},
  publisher = {AIP Publishing},
  doi     = {10.1063/5.0219552}
}

@book{nrl:2002:huba,
  title     = {NRL plasma formulary},
  author    = {Huba, J. D.},
  year      = {2002},
  publisher = {Naval Research Laboratory},
  address   = {Washington, DC}
}

@article{pop:2013:macheret,
  title     = {Kinetic modeling of the Townsend breakdown in argon},
  author    = {Macheret, S. O. and Shneider, M. N.},
  journal   = {Physics of Plasmas},
  volume    = {20},
  number    = {10},
  pages     = {101608},
  year      = {2013},
  publisher = {AIP Publishing LLC},
  doi       = {10.1063/1.4823471},
  url       = {https://doi.org/10.1063/1.4823471}
}

@article{psst:1995:tsendin,
  author  = {Tsendin, L. D.},
  title   = {Electron kinetics in non-uniform glow discharge plasmas},
  journal = {Plasma Sources Science and Technology},
  volume  = {4},
  number  = {2},
  pages   = {200--211},
  year    = {1995},
  doi     = {10.1088/0963-0252/4/2/004}
}

@article{psst:1999:phelps,
  author    = {Phelps, A. V. and Petrovi\'{c}, Z. Lj.},
  title     = {Cold-cathode discharges and breakdown in argon: surface and gas phase production of secondary electrons},
  journal   = {Plasma Sources Science and Technology},
  volume    = {8},
  number    = {3},
  pages     = {R21--R44},
  year      = {1999},
  publisher = {IOP Publishing Ltd},
  doi     = {10.1088/0963-0252/8/3/201}
}

@techreport{report:2003:hayashi,
  author      = {Hayashi, M.},
  title       = {Bibliography of electron and photon cross sections with atoms and molecules published in the 20th century - argon},
  institution = {National Institute for Fusion Science of Japan},
  number      = {NIFS-DAT-72},
  year        = {2003}
}

@article{pr:1954:bernstein,
  author  = {Bernstein, I. B. and Holstein, T.},
  title   = {Electron Energy Distributions in Stationary Discharges},
  journal = {Physical Review},
  volume  = {94},
  number  = {6},
  pages   = {1475--1482},
  year    = {1954},
  doi     = {10.1103/PhysRev.94.1475}
}

@article{pra:1987:phelps,
  author  = {Phelps, A. V. and Jelenkovi\'c, B. M. and Pitchford, L. C.},
  title   = {Simplified models of electron excitation and ionization at very high {E/n}},
  journal = {Physical Review A},
  volume  = {36}, number = {11}, pages = {5327--5336}, year = {1987},
  doi     = {10.1103/PhysRevA.36.5327}
}

@article{jpd:1987:carman,
  author  = {Carman, R. J. and Maitland, A.},
  title   = {A simulation of electron motion in the cathode sheath region of a glow discharge in helium},
  journal = {Journal of Physics D: Applied Physics},
  volume  = {20}, pages = {1021--1030}, year = {1987}
}

@article{jpd:2009:derzsi,
  author  = {Derzsi, A. and Hartmann, P. and Korolov, I. and Kar\'acsony, J. and B\'an\'o, G. and Donk\'o, Z.},
  title   = {On the accuracy and limitations of fluid models of the cathode region of dc glow discharges},
  journal = {Journal of Physics D: Applied Physics},
  volume  = {42}, pages = {225204}, year = {2009},
  doi     = {10.1088/0022-3727/42/22/225204}
}

@article{tpl:2003:tkachev,
  author  = {Tkachev, A. N. and Yakovlenko, S. I.},
  title   = {The mechanism of electron runaway in a gas and a criterion of the self-sustained discharge initiation},
  journal = {Technical Physics Letters},
  volume  = {29}, number = {8}, pages = {683--686}, year = {2003},
  doi     = {10.1134/1.1606788}
}

@article{jcp:1979:vanleer,
        author = "Van Leer, B.",
        TITLE = "Towards the Ultimate Conservative Difference Scheme.
          V. A second-Order Sequel to Godunov's Method",
        journal = "Journal of Computational Physics",
        volume = 32,
        pages = "101--136",
        year = 1979,
        doi = "10.1016/0021-9991(79)90145-1"
        }

@article{misc:1991:raizer,
  title={Longitudinal structure of the cathode portions of glow discharges},
  author={Raizer, Yurii Petrovich and Shneider, Mikhail Naumovich},
  journal={Teplofizika vysokikh temperatur},
  volume={29},
  number={6},
  pages={1041--1052},
  year={1991},
  publisher={Russian Academy of Sciences, Branch of Power Industry, Machine Building~…}
}

@article{misc:1989:raizer,
  title={Simplified kinetic equation for electrons in nonuniform fields of arbitrary strength in connection with the cathode sheath of a glow discharge},
  author={Raizer, Yu P and Shneider, MN},
  journal={Sov. J. Plasma Phys},
  volume={15},
  number={3},
  pages={184--189},
  year={1989}
}

@article{apl:1990:pak,
  author  = {Pak, H. and Kushner, M. J.},
  title   = {Multi-beam-bulk model for electron transport during commutation in an optically triggered pseudospark thyratron},
  journal = {Applied Physics Letters},
  volume  = {57},
  number  = {16},
  pages   = {1619--1621},
  year    = {1990},
  doi     = {10.1063/1.104066}
}

@article{jpd:1991:kulikovsky,
  author  = {Kulikovsky, A. A.},
  title   = {Hydrodynamic description of electron multiplication in the cathode region: elementary beams model},
  journal = {Journal of Physics D: Applied Physics},
  volume  = {24}, number = {11}, pages = {1954--1963}, year = {1991}
}

@article{ppr:2023:bochkov,
  author  = {Bochkov, E. I.},
  title   = {Multigroup Model for Treatment of Fast Electrons in Weakly Ionized Cold Plasma},
  journal = {Plasma Physics Reports},
  volume  = {49},
  number  = {2},
  pages   = {254--264},
  year    = {2023},
  doi     = {10.1134/S1063780X22601742}
}

@article{ppr:2023b:bochkov,
  author  = {Bochkov, E. I.},
  title   = {Hybrid Fluid Model for Calculating Electron Transport in Air in Strong Electric Fields},
  journal = {Plasma Physics Reports},
  volume  = {49},
  number  = {11},
  pages   = {1332--1340},
  year    = {2023},
  doi     = {10.1134/S1063780X23601311}
}

@article{ppr:2024:bochkov,
  author  = {Bochkov, E. I.},
  title   = {Multigroup Model of Calculating the Low-Energy Electron Transport in Weakly Ionized Gases},
  journal = {Plasma Physics Reports},
  volume  = {50},
  number  = {5},
  pages   = {627--640},
  year    = {2024},
  doi     = {10.1134/S1063780X24600191}
}

@article{ppr:2025:bochkov,
  author  = {Bochkov, E. I.},
  title   = {Calculating Electron Swarm Parameters in Argon in Strong Electric Fields},
  journal = {Plasma Physics Reports},
  volume  = {51},
  number  = {9},
  pages   = {1086--1099},
  year    = {2025},
  doi     = {10.1134/S1063780X25603621}
}

@article{jetp:2007:babich,
  author  = {Babich, L. P. and Kudryavtseva, M. L.},
  title   = {Group equations for moments of the relativistic electron distribution function in a cold gas of neutral atomic particles in an external electric field},
  journal = {Journal of Experimental and Theoretical Physics},
  volume  = {104},
  number  = {5},
  pages   = {704--714},
  year    = {2007},
  doi     = {10.1134/S1063776107050044}
}

@article{psst:2020:yao,
  author  = {Yao, J. and Yuan, C. and Eliseev, S. and Kudryavtsev, A. and Zhou, Z.},
  title   = {Longitudinal structure and plasma parameters of an entire {DC} glow discharge as obtained using a {1D} fluid-based model with non-local ionization},
  journal = {Plasma Sources Science and Technology},
  volume  = {29}, pages = {075003}, year = {2020},
  doi     = {10.1088/1361-6595/ab993c}
}

@article{physica:1940:kruithof,
  author  = {Kruithof, A. A.},
  title   = {Townsend's ionization coefficients for neon, argon, krypton and xenon},
  journal = {Physica},
  volume  = {7},
  number  = {6},
  pages   = {519--540},
  year    = {1940},
  doi     = {10.1016/S0031-8914(40)90043-X}
}

@article{prsa:1978:bhasavanich,
  author  = {Bhasavanich, D. and Parker, A. B.},
  title   = {The dielectric breakdown of gases at low pressure},
  journal = {Proceedings of the Royal Society of London A},
  volume  = {358},
  number  = {1694},
  pages   = {385--403},
  year    = {1978},
  doi     = {10.1098/rspa.1978.0018}
}

@article{pre:1997:petrovic,
  author  = {Petrovi\'{c}, Z. Lj. and Phelps, A. V.},
  title   = {Temporal and constriction behavior of low-pressure,
             cathode-dominated argon discharges},
  journal = {Physical Review E},
  volume  = {56},
  number  = {5},
  pages   = {5920--5931},
  year    = {1997},
  doi     = {10.1103/PhysRevE.56.5920}
}

@article{jap:1998:auday,
  author  = {Auday, G. and Guillot, Ph. and Galy, J. and Brunet, H.},
  title   = {Experimental study of the effective secondary emission coefficient
             for rare gases and copper electrodes},
  journal = {Journal of Applied Physics},
  volume  = {83},
  number  = {11},
  pages   = {5917--5921},
  year    = {1998},
  doi     = {10.1063/1.367455}
}

@article{pra:1989:gottscho,
  author  = {Gottscho, R. A. and Mitchell, A. and Scheller, G. R. and Chan, Y. and Graves, D. B.},
  title = {Electric field reversals in dc negative glow discharges},
  journal = {Physical Review A},
  volume  = {40}, pages = {6407}, year = {1989},
  doi     = {10.1103/PhysRevA.40.6407}
}

@article{psst:2024:zhou,
  author  = {Zhou, C. and Rafatov, I. and Wang, Y. and Kudryavtsev, A. and Yuan, C. and Yao, J. and Zhou, Z.},
  title   = {On fundamental inconsistencies in a commonly used modification of a fluid model for glow discharge},
  journal = {Plasma Sources Science and Technology},
  volume  = {33}, pages = {077001}, year = {2024},
  doi     = {10.1088/1361-6595/ad5ebc}
}

@article{pop:2012b:rafatov,
  author  = {Rafatov, I. and Bogdanov, E. A. and Kudryavtsev, A. A.},
  title   = {Account of nonlocal ionization by fast electrons in the fluid models of a direct current glow discharge},
  journal = {Physics of Plasmas},
  volume  = {19}, pages = {093503}, year = {2012},
  doi     = {10.1063/1.4752419}
}

@article{pop:2012:rafatov,
  author  = {Rafatov, I. and Bogdanov, E. A. and Kudryavtsev, A. A.},
  title   = {On the accuracy and reliability of different fluid models of the direct current glow discharge},
  journal = {Physics of Plasmas},
  volume  = {19}, pages = {033502}, year = {2012},
  doi     = {10.1063/1.3688875}
}

@article{pra:1990:surendra,
  author  = {Surendra, M. and Graves, D. B. and Jellum, G. M.},
  title   = {Self-consistent model of a direct-current glow discharge: Treatment of fast electrons},
  journal = {Physical Review A},
  volume  = {41}, number = {2}, pages = {1112--1125}, year = {1990},
  doi     = {10.1103/PhysRevA.41.1112}
}

@article{zp:1962:muller,
  author  = {M\"uller, K. G.},
  title = {Die Elektronenbewegung im Neutralgas bei hohen elektrischen Feldern},
  journal = {Zeitschrift f\"ur Physik},
  volume  = {169}, pages = {432--455}, year = {1962},
  doi     = {10.1007/BF01378171}
}

@article{pra:1992:kolobov,
  author  = {Kolobov, V. I. and Tsendin, L. D.},
  title   = {Analytic model of the cathode region of a short glow discharge in light gases},
  journal = {Physical Review A},
  volume  = {46},
  number  = {12},
  pages   = {7837--7852},
  year    = {1992},
  doi     = {10.1103/PhysRevA.46.7837}
}

@misc{lxcat:hayashi,
  author       = {{LXCat}},
  title        = {{Hayashi database}},
  howpublished = {www.lxcat.net},
  note         = {Retrieved on May 16, 2026}
}

@article{Belenguer1990,
  author  = {Belenguer, Ph. and Boeuf, J. P.},
  title   = {Transition between different regimes of rf glow discharges},
  journal = {Physical Review A},
  year    = {1990},
  volume  = {41},
  number  = {8},
  pages   = {4447--4459},
  doi     = {10.1103/physreva.41.4447}
}

@article{Sommerer1992,
  author  = {Sommerer, Timothy J. and Kushner, Mark J.},
  title   = {Numerical investigation of the kinetics and chemistry of rf glow discharge plasmas sustained in He, N2, O2, He/N2/O2, He/CF4/O2, and SiH4/NH3 using a Monte Carlo-fluid hybrid model},
  journal = {Journal of Applied Physics},
  year    = {1992},
  volume  = {71},
  pages   = {1654--1673},
  doi     = {10.1063/1.351196}
}

@article{Becker_2017,
  title = {Advanced fluid modeling and {PIC/MCC} simulations of low-pressure ccrf discharges},
  author = {Becker, M M and K{\"a}hlert, H and Sun, A and Bonitz, M and Loffhagen, D},
  journal = {Plasma Sources Science and Technology},
  volume = {26},
  number = {4},
  pages = {044001},
  year = {2017},
  publisher = {{IOP} Publishing},
  doi = {10.1088/1361-6595/aa5cce},
  url = {https://doi.org/10.1088/1361-6595/aa5cce}
}

@article{Schulze_2011,
  author = {Schulze, J and Donk{\'o}, Z and Sch{\"u}ngel, E and Czarnetzki, U},
  title = {Secondary electrons in dual-frequency capacitive radio frequency discharges},
  journal = {Plasma Sources Science and Technology},
  volume = {20},
  number = {4},
  pages = {045007},
  year = {2011},
  publisher = {IOP Publishing},
  doi = {10.1088/0963-0252/20/4/045007},
  url = {https://doi.org/10.1088/0963-0252/20/4/045007}
}

@article{Wilczek_2015,
  author = {Wilczek, S and Trieschmann, J and Schulze, J and Schuengel, E and Brinkmann, R P and Derzsi, A and Korolov, I and Donk{\'o}, Z and Mussenbrock, T},
  title = {The effect of the driving frequency on the confinement of beam electrons and plasma density in low-pressure capacitive discharges},
  journal = {Plasma Sources Science and Technology},
  volume = {24},
  number = {2},
  pages = {024002},
  year = {2015},
  publisher = {{IOP} Publishing},
  doi = {10.1088/0963-0252/24/2/024002},
  url = {https://doi.org/10.1088/0963-0252/24/2/024002}
}

@article{Kortshagen_1996,
  title = {On simplifying approaches to the solution of the Boltzmann equation in spatially inhomogeneous plasmas},
  author = {Kortshagen, U and Busch, C and Tsendin, L D},
  journal = {Plasma Sources Science and Technology},
  volume = {5},
  pages = {1--17},
  year = {1996},
  publisher = {IOP Publishing Ltd},
  doi     = {10.1088/0963-0252/5/1/001}
}

@article{Zhang_2024,
  author = {Zhang, Ze-Hui and Zhong, Ke-Xin and Liu, Yue and Wang, Wei and Wang, Yi-Nan and Yang, De-Zheng},
  title = {Fluid simulation of atmospheric argon RF dielectric barrier discharges: Role of neutral gas temperature},
  journal = {Physics of Plasmas},
  volume = {31},
  number = {5},
  pages = {053515},
  year = {2024},
  publisher = {AIP Publishing},
  doi = {10.1063/5.0202078},
  url = {https://doi.org/10.1063/5.0202078}
}

@article{Kortshagen_1995,
  title = {Experimental investigation and fast two-dimensional self-consistent kinetic modeling of a low-pressure inductively coupled rf discharge},
  author = {Kortshagen, U. and Pukropski, I. and Tsendin, L. D.},
  journal = {Physical Review E},
  volume = {51},
  number = {6},
  pages = {6063--6078},
  year = {1995},
  publisher = {The American Physical Society},
  doi     = {10.1103/PhysRevE.51.6063}
}

@article{Bogaerts_1999,
  author = {Bogaerts, Annemie and Gijbels, Renaat and Goedheer, Wim},
  title = {Hybrid Modeling of a Capacitively Coupled Radio Frequency Glow Discharge in Argon: Combined Monte Carlo and Fluid Model},
  journal = {Japanese Journal of Applied Physics},
  volume = {38},
  number = {7B},
  pages = {4404--4415},
  year = {1999},
  publisher = {Publication Board, Japanese Journal of Applied Physics},
  doi     = {10.1143/JJAP.38.4404}
}

@article{Sommerer_1993,
  author = {Sommerer, Timothy J},
  title = {Modelling electron kinetics in non-uniform electric fields},
  journal = {Plasma Sources Science and Technology},
  volume = {2},
  number = {3},
  pages = {198--203},
  year = {1993},
  publisher = {IOP Publishing Ltd},
  doi = {10.1088/0963-0252/2/3/007},
  url = {https://doi.org/10.1088/0963-0252/2/3/007}
}

@article{Hashiguchi_1993,
  author = {Hashiguchi, Seishiro},
  title = {Numerical Simulations of {DC} Glow Discharge using Self-Consistent Beam Model},
  journal = {Japanese Journal of Applied Physics},
  volume = {32},
  number = {6A},
  pages = {2865--2872},
  year = {1993},
  publisher = {The Japan Society of Applied Physics},
  doi     = {10.1143/JJAP.32.2865}
}

@article{Quraishi_2020,
  author = {Quraishi, Arsad and Hamo, Omri and Kronhaus, Igal},
  title = {Plasma Simulation Benchmark for {CCRF} Discharge With Secondary Electron Emission},
  journal = {IEEE Transactions on Plasma Science},
  volume = {48},
  number = {1},
  pages = {109--118},
  month = {January},
  year = {2020},
  publisher = {IEEE},
  doi = {10.1109/TPS.2019.2958972}
}

@article{Zhao_2018,
  title = {Numerical study on discharge characteristics influenced by secondary electron emission in capacitive RF argon glow discharges by fluid modeling},
  author = {Zhao, Lu-Lu and Liu, Yue and Samir, Tagra},
  journal = {Chinese Physics B},
  volume = {27},
  number = {2},
  pages = {025201},
  year = {2018},
  publisher = {Chinese Physical Society and IOP Publishing Ltd},
  doi = {10.1088/1674-1056/27/2/025201}
}

@article{Young_1993,
  author = {Young, Fongray Frank and Wu, Chwan-Hwa},
  title = {Two-Dimensional, Self-Consistent, Three-Moment Simulation of {RF} Glow Discharges},
  journal = {IEEE Transactions on Plasma Science},
  volume = {21},
  number = {3},
  pages = {312--321},
  month = {June},
  year = {1993},
  publisher = {IEEE}
}

@article{Eremin_2015,
  author = {Eremin, Denis and Hemke, Torben and Mussenbrock, Thomas},
  title = {A new hybrid scheme for simulations of highly collisional {RF}-driven plasmas},
  journal = {Plasma Sources Science and Technology},
  volume = {25},
  number = {1},
  pages = {015009},
  year = {2015},
  month = {December},
  publisher = {{IOP} Publishing},
  doi = {10.1088/0963-0252/25/1/015009}
}

@article{Wen_2022,
  author = {Wen, Ying-Ying and Li, Xin-Yang and Zhang, Yu-Ru and Song, Yuan-Hong and Wang, You-Nian},
  title = {Electron power absorption mode transition in capacitively coupled Ar/$CF_{4}$ discharges: hybrid modeling investigation},
  journal = {Journal of Physics D: Applied Physics},
  volume = {55},
  number = {20},
  pages = {200001},
  month = {February},
  year = {2022},
  publisher = {IOP Publishing},
  doi = {10.1088/1361-6463/ac52cd},
  url = {https://doi.org/10.1088/1361-6463/ac52cd}
}

@article{Noesges_2023,
  author = {Noesges, K and Klich, M and Derzsi, A and Horv{\'{a}}th, B and Schulze, J and Brinkmann, R P and Mussenbrock, T and Wilczek, S},
  title = {Nonlocal dynamics of secondary electrons in capacitively coupled radio frequency discharges},
  journal = {Plasma Sources Science and Technology},
  volume = {32},
  number = {8},
  pages = {085008},
  year = {2023},
  publisher = {IOP Publishing},
  doi = {10.1088/1361-6595/ace848},
  url = {https://doi.org/10.1088/1361-6595/ace848}
}

@article{Derzsi_2019,
  author = {Derzsi, Aranka and Horváth, Bence and Korolov, Ihor and Donkó, Zoltán and Schulze, Julian},
  title = {Heavy-particle induced secondary electrons in capacitive radio frequency discharges driven by tailored voltage waveforms},
  journal = {Journal of Applied Physics},
  volume = {126},
  number = {4},
  pages = {043303},
  year = {2019},
  publisher = {AIP Publishing},
  doi = {10.1063/1.5100508},
  url = {https://doi.org/10.1063/1.5100508}
}

@article{Chanrion_2014,
  author = {Chanrion, O and Bonaventura, Z and Çinar, D and Bourdon, A and Neubert, T},
  title = {Runaway electrons from a 'beam-bulk' model of streamer: application to {TGFs}},
  journal = {Environmental Research Letters},
  volume = {9},
  number = {5},
  pages = {055003},
  month = {May},
  year = {2014},
  publisher = {IOP Publishing},
  doi = {10.1088/1748-9326/9/5/055003}
}
\bibliographystyle{aiaa2}
\end{document}